\documentclass{article}
\usepackage{arxiv}
\usepackage[utf8]{inputenc} 
\usepackage{graphicx}
\usepackage{subcaption}
\usepackage[T1]{fontenc}    
\usepackage{hyperref}       
\usepackage{url}            
\usepackage{booktabs}       
\usepackage{amsfonts}       
\usepackage{nicefrac}       
\usepackage{microtype}      
\usepackage{lipsum}
\usepackage[table]{xcolor}
\usepackage{amsmath,amssymb}

\usepackage{multirow}

\usepackage{subcaption}
\usepackage{diagbox}
\usepackage{hhline}

\definecolor{lightgray}{gray}{0.9}

\newcolumntype{+}{!{\vrule width 2pt}}

\newlength\savedwidth

\newcommand\thickhline{\noalign{\global\savedwidth\arrayrulewidth\global\arrayrulewidth 2pt}%
\hline
\noalign{\global\arrayrulewidth\savedwidth}}

\begin{document}

\title{Time evolution of the hierarchical networks between PubMed MeSH terms}
\author{
  S\'{a}muel G. Balogh\\
  Dept. of Biological Physics\\ 
  E\"{o}tv\"{o}s University\\
  H-1117 Budapest, Hungary\\
  \texttt{balogh@hal.elte.hu} \\
   \And
    D\'{a}niel Zagyva\\
  Dept. of Biological Physics\\ 
  E\"{o}tv\"{o}s University\\
  H-1117 Budapest, Hungary\\
   \AND
 P\'{e}ter Pollner \\
  MTA-ELTE Statistical and Biological Physics Research Group\\
  Hungarian Academy of Sciences\\
  H-1117 Budapest, Hungary\\
   \And
   Gergely Palla \\
  MTA-ELTE Statistical and Biological Physics Research Group\\
  Hungarian Academy of Sciences\\
  H-1117 Budapest, Hungary\\
}


\begin{abstract}
Hierarchical organisation is a prevalent feature of many complex networks appearing in nature and society. A relating interesting, yet less studied question is how does a hierarchical network evolve over time? Here we take a data driven approach and examine the time evolution of the network between the Medical Subject Headings (MeSH) provided by the National Center for Biotechnology Information (NCBI, part of the U. S. National Library of Medicine). The network between the MeSH terms is organised into 16 different, yearly updated hierarchies such as ``Anatomy'', ``Diseases'', ``Chemicals and Drugs'', etc. The natural representation of these hierarchies is given by directed acyclic graphs, composed of links pointing from nodes higher in the hierarchy towards nodes in lower levels.  Due to the yearly updates, the structure of these networks is subject to constant evolution: new MesH terms can appear, terms becoming obsolete can be deleted or be merged with other terms, and also already existing parts of the network may be rewired. We examine various statistical properties of the time evolution, with a special focus on the attachment and detachment mechanisms of the links, and find a few general features that are characteristic for all MesH hierarchies. According to the results, the hierarchies investigated display an interesting interplay between non-uniform preference with respect to multiple different topological and hierarchical properties.
\end{abstract}

\flushbottom
\maketitle

\thispagestyle{empty}

\section*{Introduction}

In the recent decades the network approach has become fundamental in the studies of various phenomena in nature and society, ranging from the level of interactions within cells to the level of the Internet, economic networks, and the society \cite{Laci_revmod,Dorog_book}. A very important topic in this field is related to the hierarchical organization of networks \cite{Laci_hier_scale,Newman_hier,Pumain_book,Sole_chaos_hier,Anna_and_Tamas_book}. Grasping the signs of hierarchy in networks is a non-trivial task with a number of possible different approaches, including the statistical inference of an underlying hierarchy based on the observed network structure \cite{Newman_hier}, and the introduction of various hierarchy measures \cite{Sneppen_hier_measures,Enys_hierarchy,Sole_hier_PNAS,RWH,Gupte_hier_measure,Elisa_hier_measure}. Examples of empirical studies on hierarchical networks are including the transcriptional regulatory network of Escherichia coli \cite{Zeng_Ecoli}, the dominant-subordinate hierarchy among crayfish \cite{Huber_crayfish}, the leader-follower network of pigeon flocks \cite{Tamas_pigeons,Pigeon_context} and harems of Przewalski horses \cite{Ozogany}, the rhesus macaque kingdoms \cite{McCowan_macaque}, neural networks \cite{Kaiser_neural} and technological networks \cite{Pumain_book}, scientific journals \cite{Palgrave}, social interactions \cite{Guimera_hier_soc,our_pref_coms,Sole_hier_soc,PicturAsk}, urban planning \cite{Krugman_urban,Batty_urban}, on-line news content \cite{news_portals}, ecological systems \cite{Hirata_eco,Wickens_eco}, and evolution \cite{Eldrege_book,McShea_organism,Mengistu_evolv_hier}. In addition, hierarchical organisation is also related to the non-normality of networks \cite{Lambiotte_non_normal}, the topological properties of various scientific and techno-scientific fields \cite{Palgrave,Katchanov2017,science_mapping,Chen_book_chap,graphene_paper,sustainability_paper} (usually depicted by citation networks), and the optimal performance of interacting agent groups \cite{Anna_Tamas_Nature_Communications,Zamani_2017,Zamani_2018}. Hierarchies are usually depicted as directed acyclic graphs, in which the links are not allowed to form directed cycles, and where a pair of nodes connected by a link are assumed to be in some sort of asymmetric relationship with each other such as parents and children, leaders and followers, etc.

Networks representing real systems are subject to constant evolution in most of the cases, and some relevant aspects of the laws forming the shape of networks changing over time have already been uncovered in the scientific literature. Probably most famous is the preferential attachment rule for growing scale-free networks, which is one of the key concepts of the Barabási-Albert model \cite{BA_model}, and was detected also by empirical studies of network data \cite{Tamas-mer,Barab-mer,Newman-mer,pref-coms}. Another notable example is provided by the studies of the various statistical features of community evolution in networks \cite{group_evolv_nature}. Along the same line, in the present paper our aim is to examine the statistical properties of time dependent networks with a hierarchical structure. 

Our study is based on the data provided by the NCBI about the MeSH terms, which were introduced for helping the search in the PubMed publication database of the NCBI (comprising more than 26 million citations for biomedical literature) at various levels of specificity. The MeSH terms are hierarchically organized: At the most general level of the hierarchical structure we find very broad headings such as ``Organisms'' or ``Information Science'', whereas more specific headings are found at deeper (more narrow) levels. Due to the rapidly developing nature of the medical-, biochemical- and  biological sciences, the set of available MeSH terms are yearly updated by the administrators of PubMed. This provides a fascinating empirical data-set for the study of time dependent hierarchical networks. A few previous studies on this data-set have already been published, approaching the development of the Mesh term hierarchies from an ontological perspective \cite{Tsatsaronis,Leengheer,McCray,Tsatsaronis_2}. The main focus of these results was on the growth of the system, concentrating on how are the newly introduced Mesh terms categorized and linked under already existing older Mesh terms. In our present study we show that restructuring plays an equally important role in forming the structure of the Mesh hierarchies. 

Our goal is to examine the statistical features of the time evolution in the observed hierarchies. One of the central questions we are interested in is how do the different topological- and hierarchical properties of the nodes influence the attachment and detachment of links during the restructuring. Understanding the nature of these processes can help the creation of hierarchy evolution models that can predict which part of the hierarchy is most likely to be rewired in the future, and what is the expected change in the overall features of the hierarchy.

\section*{Data and methods}
\subsection*{Basic properties of the MeSH hierarchies}
The directed networks we consider are based on the classifications provided by PubMed, specifying at least one parent for any available MesH term, except for the roots of the hierarchies. The raw data we use is publicly available on the link provided in Ref. \cite{raw_MeSH_data}. There are altogether 16 different roots, and the total number of descendants of the individual roots (the sizes of the hierarchies) varies roughly between a 1,00 and a 10,000 nodes, whereas the time span of our analysis is 14 years. In Table.\ref{table:basic_props}. we list a few basic properties of these networks, including the minimum and maximum sizes, the maximum level depth and the average fraction of changed links under one time step (one year). In the Supporting Information we also provide more detailed tables listing the yearly size of the hierarchies, together with the number of added and deleted nodes and links.

\begin{table}[!ht]
\rowcolors{1}{white}{lightgray}
\centering
\caption{
{\bf Basic hierarchy data}}
\begin{tabular}{|l+l|c|c|c|}
\hline
\multirow{2}{*}{\cellcolor{white}}   &  &  & max. & 
average \\
\cellcolor{white} & \multirow{-2}{*}{\cellcolor{white} Root name}  & 
\multirow{-2}{*}{\cellcolor{white} size range}  & \cellcolor{white} depth &
\cellcolor{white} change  \\ \thickhline
\rowcolor{lightgray}
A & Anatomy & 1350 - 1826 & 10 & 4.66\% \\
\hline
\rowcolor{white}B & Organisms &  2252 - 3815  & 13   & 6.49\%  \\
\hline
\rowcolor{lightgray}C & Diseases  &  3975 - 4799 & 8 & 4.23\% \\
\hline
\rowcolor{white}D & Chemicals and Drugs & 6902 - 9934 & 11 & 6.22\% \\
\hline
\rowcolor{lightgray} & Analytical, Diagnostic and &  &  &  \\
 \multirow{-2}{*}{E}& Therapeutic Techniques and Equipment &
\multirow{-2}{*}{2040 - 2924}  & \multirow{-2}{*}{9} &
 \multirow{-2}{*}{4.89\%} \\
\hline
F & Psychiatry and Psychology & 807 - 1083 & 7 & 3.60\% \\
\hline
G & Phenomena and Processes & 1733 - 2259 & 10 & 15.18\% \\
\hline
H & Disciplines and Occupations & 334 - 537 & 8 & 12.07\% \\
\hline
 & Anthropology, Education, Sociology &  & & \\
\rowcolor{lightgray} \multirow{-2}{*}{I}&  and Social Phenomena & \multirow{-2}{*}{449 - 641}&
\multirow{-2}{*}{9}  & \multirow{-2}{*}{5.23\%}\\
\hline
\rowcolor{white}J & Technology, Industry, Agriculture & 254 - 582 & 10 & 8.92\% \\
\hline
\rowcolor{lightgray}K & Humanities & 152 - 200 & 7 & 3.93\% \\
\hline
\rowcolor{white}L & Information Science & 322 - 476 & 9 & 5.82\% \\
\hline
\rowcolor{lightgray}M & Named Groups & 174 - 290 & 7 & 5.71\% \\
\hline 
\rowcolor{white}N & Health Care &  1072 - 1795 & 10 & 4.94\% \\
\hline
\rowcolor{lightgray}V & Publication Characteristics & 137 - 163 & 6 & 3.44\% \\
\hline
\rowcolor{white}Z & Geographicals & 369 - 402 & 6 & 1.94\% \\
\hline
\end{tabular}
\begin{flushleft} The 1$^{\rm st}$ column lists the hierarchy ID, the 2$^{\rm nd}$ gives the name of the root, the 3$^{\rm rd}$ column provides the minimum and maximum sizes during the time evolution, the 4$^{\rm th}$ contains the maximum level depth, and finally the 5$^{\rm th}$ column lists the average fraction of changed links under one year. 
\end{flushleft}
\label{table:basic_props}
\end{table}

\begin{figure}[h]
\centering
\includegraphics[width=0.9\textwidth]{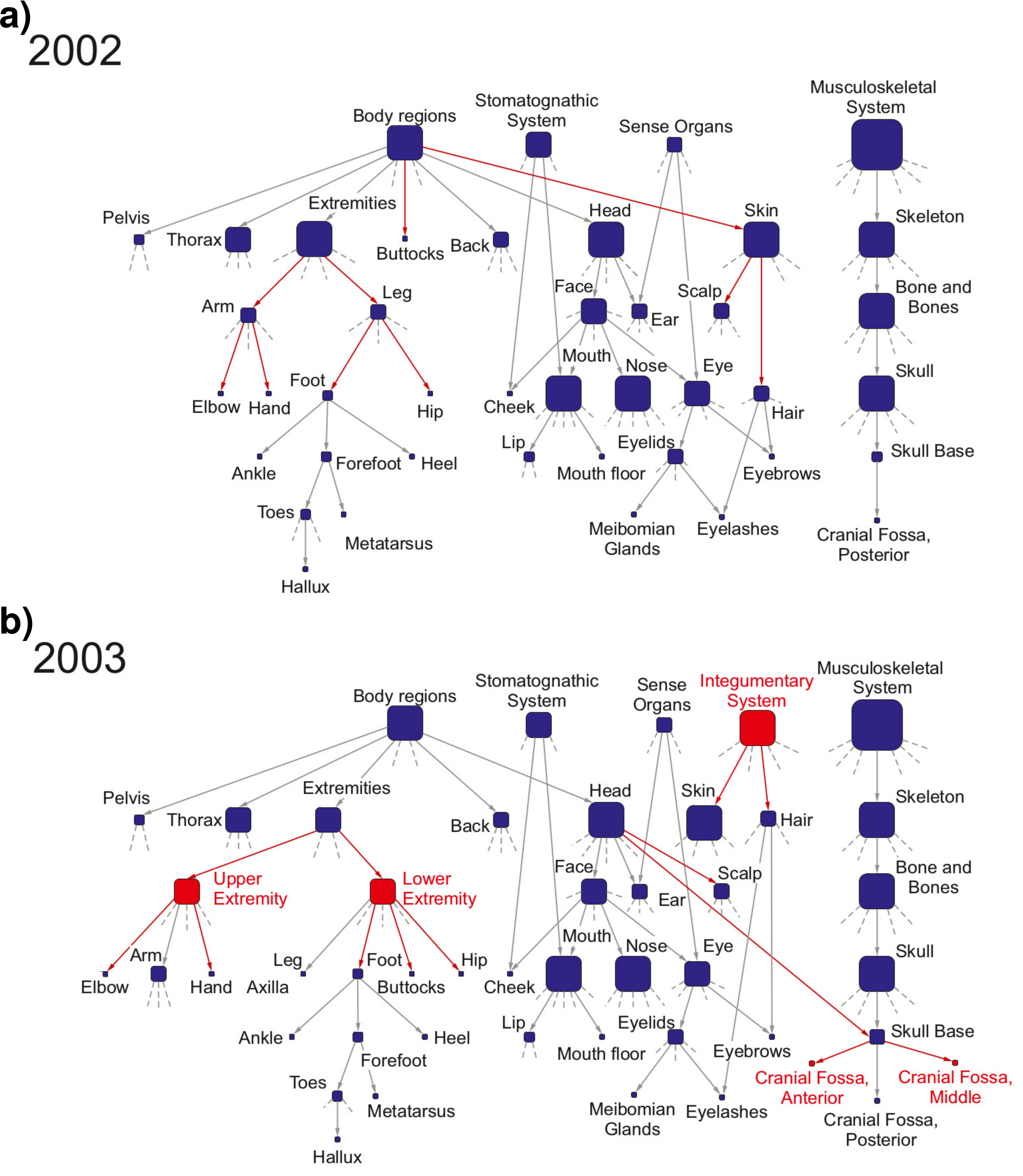}
\caption{{\bf Changes between subsequent time steps in a MeSH hierarchy} a) A small part of the hierarchy 'A' (Anatomy) in 2002. Red links are deleted in the next time step b) The corresponding part of the same hierarchy in 2003. Nodes and links colored red are newly appearing elements.}
\label{fig:Fig1}
\end{figure}

The links in our network representation are pointing from the parents to their children. Since a part of the MesH terms have multiple parents, the studied networks are not strictly tree-like, instead they correspond to a directed acyclic graphs. Due to the yearly updates, the structure of these networks is subject to constant evolution: New MesH terms can appear, terms becoming obsolete can be deleted or be merged with other terms, and also already existing parts of the network may be rewired. To illustrate these processes, in Fig.\ref{fig:Fig1}. we show two snapshots from subsequent years, depicting the changes in a small subgraph from the hierarchy A (Anatomy).   
According to the picture, a relatively large variety of modifications can occur already in a single time step. 
E.g.,  'Cranial Fossa Anterior' is a newly appearing MeSH term, which is classified under 'Skull Base' in Fig.\ref{fig:Fig1}b. This type of process can be viewed in general as the growth of the hierarchy. Another intuitive process is rewiring,  when both the source and the target of a newly appearing link are actually already existing ('old') nodes, such as e.g., the new link between 'Head' and 'Scalp' in Fig.\ref{fig:Fig1}b. Naturally, links becoming obsolete can also become deleted, as e.g., the link from 'Body regions' to 'Skin' in Fig.\ref{fig:Fig1}a. There are also somewhat less intuitive change types as well, such as the insertion of a new node into the middle part of a branch, as e.g., the link from 'Upper Extremity' to 'Arm' in Fig.\ref{fig:Fig1}b, or the appearance of a new link between two new nodes. A detailed classification of the possible change types is given in the Results section.

\subsection*{Measuring preference during attachment or detachment}

Our main focus in this paper is on the examination of possible preference with respect to various node properties during the attachment and detachment of the links. The method we use for detecting whether the attachments/detachments are uniform with respect to a given property $x$, or instead show preference towards high (or low) values of $x$ is based on comparing the distribution of $x$ for the chosen nodes during the change event and the distribution of $x$ amongst the available nodes \cite{our_pref_coms}.

\subsubsection*{Attachment events}
We begin by discussing attachment events, where (previously non existing) new links appear in the system. For simplicity let us consider first only two consecutive time steps in the data set for a single hierarchy, where we would like to examine whether the choice of nodes in the initial state is preferential or not with regard to  $x$.  We denote the probability distribution of $x$ at the initial state by $p(x)$, and the complementary cumulative distribution by $Q(x)=\sum\limits_{x'\geq x}p(x')$, corresponding to the fraction of nodes in the hierarchy having a property value at least as large as $x$ in the initial state. In case the attachment is independent of $x$, the number of nodes chosen having a property value $x$ or larger is expected to be simply proportional to $Q(x)$. However, if larger values of $x$ are preferred, then nodes having large $x$ value are chosen at a higher frequency compared to what we would expect based on $Q(x)$, and similarly, if lower values of $x$ are preferred, then nodes with large $x$ values are chosen at a lower frequency compared to the expectation based on $Q(x)$. Therefore, by denoting the number of actually chosen nodes having a property value at least as large as $x$ in the attachments  as $w(x)$, and taking its ratio compared to $Q(x)$ as
\begin{equation}
W(x) = \frac{w(x)}{Q(x)}, 
\label{eq:W_def}
\end{equation}
we obtain a function that is constant if the attachment is uniform in $x$, since in this case $w(x)$ and $Q(x)$ are simply proportional to each other for any $x$. However, if larger values of $x$ are preferred, the shape of $W(x)$ becomes increasing as a function of $x$, whereas in the opposite case, when the attachment/detachment prefers lower values of $x$, the shape of $W(x)$ becomes decreasing.

A noteworthy property of $w(x)$ is that for any fixed value of $x$, it follows a binomial distribution,
\begin{equation}
    P(w(x) =k) = {A \choose k} u(x)^k(1 - u(x))^{A-k},
    \label{eq:binom}
\end{equation}
where $A$ is the number of attachment events, and $u(x)$ denotes the probability for choosing a node having a property value at least as large as $x$. Simplest case is when choosing is independent of the given property, and therefore, $u(x)=Q(x)$. If instead we assume a linear preference with regard to the studied property, $u(x)$ can be expressed as
\begin{equation}
    u(x) = 
    \sum\limits_{i:\, x_i\geq x}x_i p(x_i)
    \bigg/\sum\limits_{j=1}^{A}x_j p(x_j),
\end{equation}
where the summations run over the nodes in the hierarchy. 

In any case, based on (\ref{eq:binom}) the expected value and standard deviation of $w(x)$ can be given as $\left< w(x)\right> = A\cdot u(x)$ and $\sigma(w(x))=\sqrt{A\cdot u(x)(1-u(x))}$, respectively. By moving from $w(x)$ to $W(x)$  we obtain that according to (\ref{eq:W_def}) the mean and standard deviation for $W(x)$ can be written as
\begin{align}
    \left< W(x)\right>&= \frac{A\cdot u(x)}{Q(x)}, \\
    \sigma\left(W(x)\right)&= \frac{\sqrt{A\cdot u(x)(1-u(x))}}{Q(x)},
\end{align}
which for an attachment process independent of $x$ take the simple form of
\begin{align}
    \left< W(x)\right>&= A, \label{eq:rand_mean}\\
    \sigma\left(W(x)\right)&= \frac{\sqrt{A(1-Q(x))}}{\sqrt{Q(x)}}. \label{eq:rand_std}
\end{align}

\begin{figure}[h]
\centering
\includegraphics[width=0.6\textwidth]{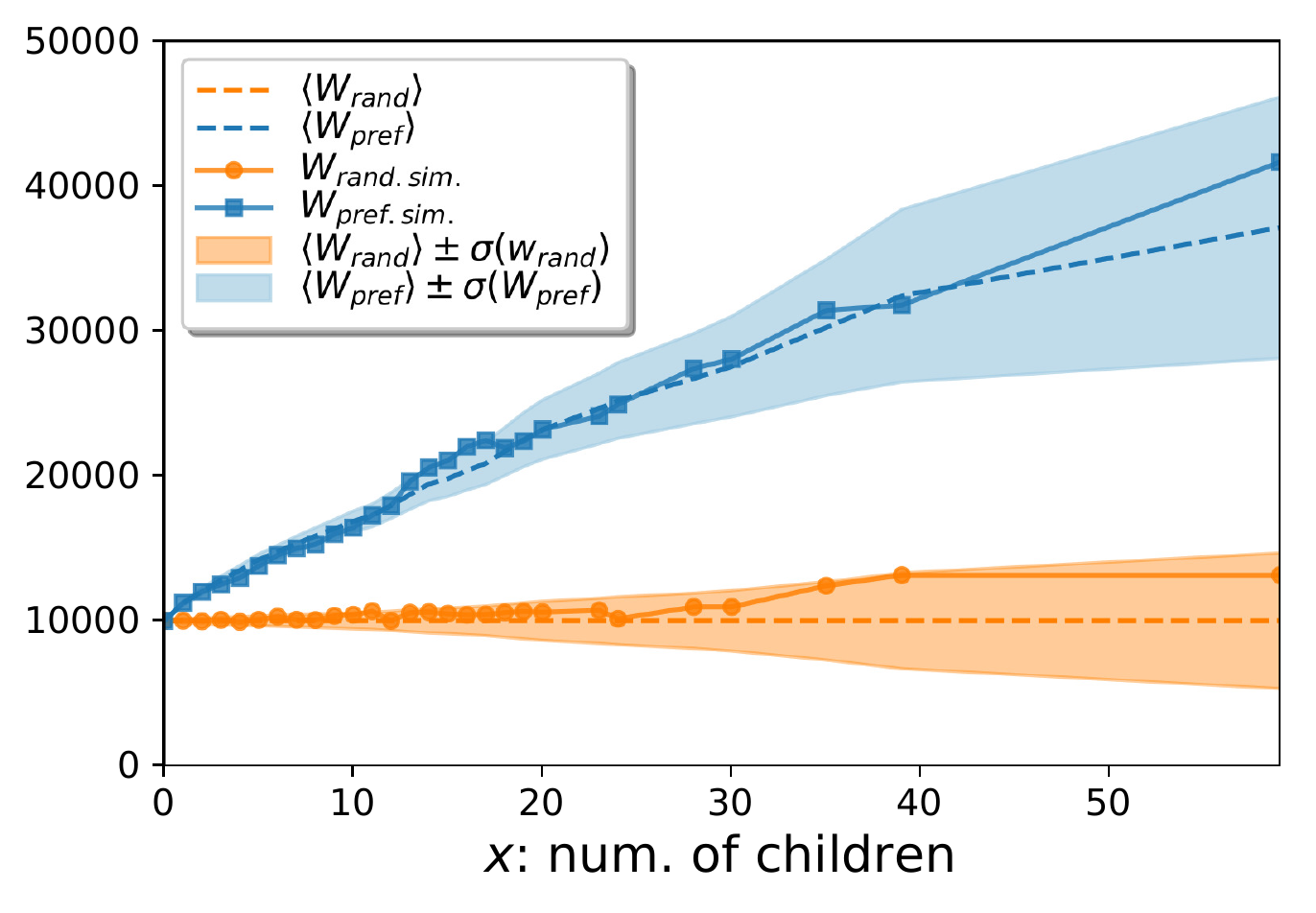}
\caption{{\bf Testing $W(x)$ by simulated attachments.} 
The property $x$ here corresponds to the number of children, and the full symbols connected by continuous lines show the measured $W(x)$ for random attachment (independent of $x$) in orange (circles), and for preferential attachment with an additive constant (i.e. when a newly added node connects to node $i$ with a probability $\frac{k_i+a}{\sum_{i}k_i+a}$ where $a$ is an arbitrary constant) in blue (squares). Dashed lines correspond to the analytic mean for $W(x)$, whereas the shaded areas indicate the standard deviation around the mean.}
\label{fig:Fig2}
\end{figure}

We have tested the behavior of $W(x)$ by simulating $A=10000$ attachment events on hierarchy $D$ at year $2002$, the results are shown in Fig.\ref{fig:Fig2}. According to the plots, the measured $W(x)$ remained within the standard deviation around the analytically calculated average for both purely random attachments (orange color), and linear preferential attachment with an additive constant. 

When applying the above method for measuring preference in the empirical data, for every time step $t$ (except for the last) we can measure the complementary cumulative distribution $Q_t(x)$, and count how many nodes having a property value at least as large as $x$ have been selected by the given attachment mechanism between $t$ and $t+1$, denoted by $w_t(x)$. By aggregating their ratio in analogy with (\ref{eq:W_def}), we can define 
\begin{equation}
    W_{\rm emp}(x) =\sum_{t=1}^{t_{\rm max}-1}\frac{w_{t}(x)}{Q_t(x)}. \label{eq:W_emp_def}
\end{equation}
This can be compared to e.g., the mean and standard deviation of the random variable corresponding to the sum of the supposed $W(x)$ under the assumption of independence from $x$, which according to (\ref{eq:rand_mean}-\ref{eq:rand_std}) can be given as
\begin{align}
    \left< W_{\rm rand}(x)\right>&= \sum\limits_{t=1}^{t_{\rm max}-1}A_t, \label{eq:rand_mean_t}\\
    \sigma\left(W_{\rm rand}(x)\right)&= \left[\sum\limits_{t=1}^{t_{\rm max}-1} \frac{A_t(1-Q_t(x))}{Q_t(x)}\right]^{\frac{1}{2}}, \label{eq:rand_std_t}
\end{align}
where $A_t$ denotes the number of attachment events between time steps $t$ and $t+1$.

\subsubsection*{Detachment events}

An important difference between the addition of new links and link deletion events is that in the latter case, the natural assumption for the random choice (independent of any node property) is choosing a link uniformly at random from all existing links. Under this process, high degree nodes appear to be involved in the link change events with higher probability compared to low degree nodes simply because they have a higher number of connection. To take this into account we have to redefine the formula for $Q(x)$.

First let us consider the case, where we are interested in whether some property $x$  has an effect on the likelihood that an out link is detached from a node (which means that from the point of view of the deleted link, the node is corresponding to the source). If we choose at random from all possible links, the probability that we pick an out link from a node with out degree $k_{\rm out}$ is given by $k_{\rm out}p(k_{\rm out})/\left< k_{\rm out}\right>$, where $p(k_{\rm out})$ denotes the out degree distribution, and $\left< k_{\rm out}\right>$ is the average out degree (which is the same as the average in degree). According to that, the probability distribution for property $x$ on the source node of randomly selected links can be written as
\begin{equation}
    p_{\rm out}(x) = \frac{1}{\left< k_{\text{out}}\right>}\sum_{k_{\rm out}} p(x \mid k_{\rm out}) k_{\rm out} p(k_{\rm out}), 
\end{equation}
where $p(x \mid k_{\rm out})$ denotes the conditional probability that the property value is $x$, given that the out degree of the node is $k_{\rm out}$. Based on $p_{\rm out}(x)$, the complementary cumulative distribution $Q_{\rm out}(x)$ can be calculated as usual, 
\begin{equation}
    Q_{\rm out}(x) = \sum_{x'\geq x} p_{\rm out}(x').
    \label{eq:Q_del_out}
\end{equation}

If in contrast to out links, we are interested in the deletion of incoming links and the possible effect on the likelihood of such events by some node property $x$, we can formulate analogous formulas to the above using the in degree distribution $p(k_{\rm in})$. In this case the probability distribution for property $x$ on the target node of randomly selected links can be written as
\begin{equation}
    p_{\rm in}(x) = \frac{1}{\left< k_{\text{in}}\right>}\sum_{k_{\rm in}} p(x \mid k_{\rm in}) k_{\rm in} p(k_{\rm in}), 
\end{equation}
where $p(x \mid k_{\rm in})$ denotes the conditional probability that the property value is $x$, given that the in degree of the node is $k_{\rm in}$, and the corresponding complementary cumulative distribution is given by
\begin{equation}
    Q_{\rm in}(x) = \sum_{x'\geq x} p_{\rm in}(x').
    \label{eq:Q_del_in}
\end{equation}

Otherwise, the analysis for the link deletion events is the same as in case of the attachment events: We can calculate $Q_t(x)$ using either (\ref{eq:Q_del_out}) or (\ref{eq:Q_del_in}), and by plugging the result together with the observed $w_t(x)$ into (\ref{eq:W_emp_def}) we obtain $W_{\rm emp}(x)$. To decide whether we can speak about a possible preference or anti-preference with respect to the chosen property, $W_{\rm emp}(x)$ has to be compared to the $W(x)$ expected based on neutral behaviour, calculated using (\ref{eq:rand_mean_t}-\ref{eq:rand_std_t}).

\section*{Results}
We applied the methodology outlined in the previous section to study the time evolution of the hierarchies listed in Table.\ref{table:basic_props}. where the system size exceeds 1000 nodes during the whole recorded time period, corresponding to hierarchies A, B, C, D, E, G, and N. Before actually showing the results, first we need to specify the different possible attachment and detachment event types. In terms of the changing links we have two large categories: added (new) links and deleted links. When examining the endpoints of added links, both the source and the target can be either an already existing (old) node, or a new node, thus, there are altogether 4 types of added links. The case of deleted links is much simpler in this respect, as both endpoints must correspond to old nodes. Therefore, there are in total 5 different possibilities for changes in the connections. However, when examining the possible effect of a given node property on the likelihood that the node is going to take part in an attachment/detachment event, we also have to specify whether the node is the source or the target of the involved link. Thus, for any node property of interest we can examine 10 different scenarios over the time evolution of the hierarchies. Naturally, when interested in the possible effect of a node property of an old node, the value of the property is always measured before the link change event (e.g., if the change occurs between time steps $t$ and $t+1$, then it is recorded at $t$), whereas for new nodes we can only measure their properties at the time point of their appearance (i.e., at $t+1$ for link change events between $t$ and $t+1$). We list the yearly frequencies of the different event types for the studied hierarchies in the Supporting Information in Tables S1-S7. 
 %
 %

\begin{figure*}[h]
\centering
\includegraphics[width=1.0\textwidth]{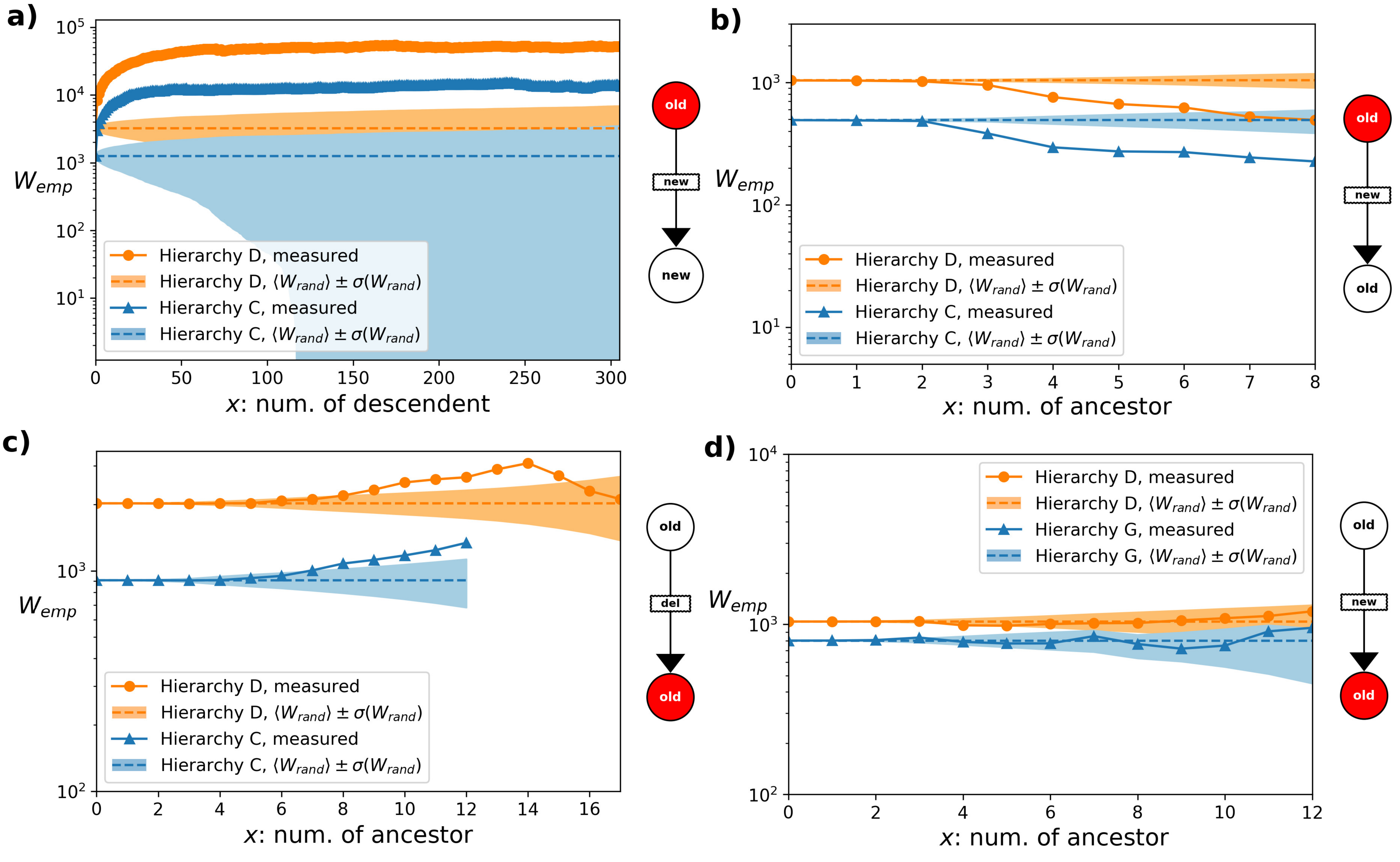}
\caption{{\bf Measuring preference in attachment and detachment events.} In each panel we compare $W_{\rm emp}(x)$ defined in (\ref{eq:W_emp_def}) to the mean and standard deviation of $W(x)$ for random events, given in (\ref{eq:rand_mean_t}-\ref{eq:rand_std_t}) and indicated by dashed lines in shaded areas. The pictograms beside the panels show the type of the studied attachment/detachment events and highlight in red whether the given property $x$ was measured on the source or on the target  of the links involved in the events. a) Results for the total number of descendants of source nodes in attachments of new links pointing from old nodes to new nodes in hierarchies D (orange) and C (blue). b) $W_{\rm emp}(x)$ for the number of ancestors of source nodes on new links appearing between old nodes, measured in hierarchies D (orange) and C (blue). c) The same plots when $x$ is equal to the number of ancestors of the target nodes in link deletion events for hierarchies D (orange) and C (blue). d) $W_{\rm emp}(x)$ in case $x$ is corresponding to the number of ancestors of the target node in attachment of new links between old nodes.} 
        \label{fig:Fig3}
\end{figure*}

In our studies we focused on the following properties: number of children (out degree), number of parents (in degree), total number of descendants, total number of ancestors. As an illustration, in Fig.\ref{fig:Fig3}. we show parts of the results obtained for hierarchies D, C and G. In Fig.\ref{fig:Fig3}a the $W_{\rm emp}(x)$ is plotted for hierarchies C and D, obtained from events where a new link pointing to a new node is attached to an old node, and $x$ is corresponding to the total number of descendants of the source node. The curves indicate strong preference for large values of $x$, as they clearly exceed $W_{\rm rand}(x)+\sigma\left(W_{\rm rand}(x)\right)$ by an order of magnitude. Interestingly somewhat the opposite can be seen in Fig.\ref{fig:Fig3}b, showing the results for the  same hierarchies in case of insertion of new links between pairs of already existing nodes, where $x$ is corresponding to the total number of ancestors of the source node. The fact that $W_{\rm emp}(x)$  is way below $W_{\rm rand}(x)-\sigma\left(W_{\rm rand}(x)\right)$ indicates that the probability for the attachment of an incoming link to a node with higher number of ancestors is lower than what we would expect at random. In Fig.\ref{fig:Fig3}c we considered link deletion events, and according to the results $W_{\rm emp}(x)$ shows a non-monotonous behaviour as a function of $x$ for the total number of ancestors of the target node in case of hierarchy D and a weak preference in case of hierarchy C. The peak in $W_{\rm emp}(x)$ for hierarchy D is suggesting that there is a preferred value of $x$, where the likelihood of the node taking part in the given type of detachment event is maximal. Finally, in Fig.\ref{fig:Fig3}d we show the results for the insertion of new links between old nodes (similarly to Fig.\ref{fig:Fig3}b), but this time we depict $W_{\rm emp}(x)$ for the total number of ancestors of the target node in case of hierarchies D and G. According to the results $W_{\rm emp}(x)$ runs within the range of the standard deviation around $\left<W_{\rm rand}(x)\right>$, thus, this type of attachment does not show any preference with respect to the number of ancestors of the target node. 

Similar plots for the rest of the attachment/detachment types and for the other hierarchies are given in the Supporting Information. Based on the seen behaviour of $W_{\rm emp}(x)$ we can categorise the observed behaviour as follows:
\begin{itemize}
    \item Strong indication of preference (s+): $W_{\rm emp}(x)$ shows a monotonous increasing behaviour, and exceeds $\left< W_{\rm rand}(x)\right> + \sigma\left(W_{\rm rand}(x)\right)$ by a large amount, (as e.g., in case of Fig.\ref{fig:Fig3}a).
    \item Weak indication of preference (w+): $W_{\rm emp}(x)$ shows a monotonous increasing behaviour, exceeds $\left< W_{\rm rand}(x)\right>+\sigma\left(W_{\rm rand}(x)\right)$, but only by a small amount.
    \item Strong indication of no preference (s0): $W_{\rm emp}(x)$ remains within the standard deviation around $\left< W(x)\right>$, (as e.g., in case of Fig.\ref{fig:Fig3}d).
    \item Weak indication of anti-preference (w-): $W_{\rm emp}(x)$ shows a monotonous decreasing behaviour, and falls under $\left< W_{\rm rand}(x)\right>-\sigma\left(W_{\rm rand}(x)\right)$, by a small amount.
    \item Strong indication of anti-preference (s-): $W_{\rm emp}(x)$ shows a monotonous decreasing behaviour, and falls under $\left< W_{\rm rand}(x)\right>-\sigma\left(W_{\rm rand}(x)\right)$, by a larger amount, (as e.g., in case of Fig.\ref{fig:Fig3}b)
    \item Indication of preference with a peak (p+): $W_{\rm emp}(x)$ shows a non-monotonous behaviour, and has a maximum exceeding $\left< W_{\rm rand}(x)\right>+\sigma\left(W_{\rm rand}(x)\right)$ by a significant amount, (as e.g., in case of Fig.\ref{fig:Fig3}c).
    \item Indication of anti-preference with a peak (p-): $W_{\rm emp}(x)$ shows a non-monotonous behaviour, and has a minimum falling under $\left< W_{\rm rand}(x)\right>-\sigma\left(W_{\rm rand}(x)\right)$ by a significant amount.
    \item Insufficient statistics (i.s): in a number of cases it is not possible to draw a conclusion based on the empirical data. This may be due to the fact that the given type of attachment/detachment occurs rarely, or because that the distribution of the given node property is extremely narrow, resulting in a very limited range for $x$. 
\end{itemize}
In Table \ref{table:hierarchy_D}. we give a summary overview of the results for the largest hierarchy (corresponding to hierarchy D), where the table is organised as follows: rows are corresponding to the 4 studied node properties, measured either on the source node (top 4 rows) or the target node (bottom 4 rows) of the changing links, and the table columns indicate the attachment/detachment types. In each cell we provide the category of the observed behaviour based on the corresponding plot. For example, the 3$^{\rm rd}$ cell in the 3$^{\rm rd}$ row is based on the orange curve in Fig.\ref{fig:Fig3}a, the 4$^{\rm th}$ cell in the 4$^{\rm th}$ row is connected to Fig.\ref{fig:Fig3}b, the 4$^{\rm th}$ cell in the last row is corresponding to Fig.\ref{fig:Fig3}d, etc. 

The overall pattern of the different preference types in Table \ref{table:hierarchy_D}. is highly non-trivial. E.g., all possible link change event types show preference with respect to the number of children of the source node (first row in Table \ref{table:hierarchy_D}.), and all except for two (addition of new links between new nodes and deletion of old links) show anti-preference with respect to the total number of ancestors of the source node (4$^{\rm th}$ row in Table \ref{table:hierarchy_D}.)\color{black}. Interestingly, in the 3$^{\rm rd}$ row of Table \ref{table:hierarchy_D}. (corresponding to the number of descendants of the source node) both preference and anti-preference is occurring among the cells corresponding to the different link change types. Seemingly the properties of the target nodes (bottom 4 rows) have a smaller effect compared to the properties of the source nodes (top 4 rows), indicated by the higher number of cells falling into the category of evidence for no-preference (s0). Nevertheless, preference with respect to the number of parents and number of ancestors, and anti-preference with respect to the number of children and number of descendants can be seen for a couple of the link change types.

\definecolor{cyan}{rgb}{0.0, 1.0, 1.0}
\definecolor{greyishcyan}{rgb}{0.65, 0.9, 0.8}
\definecolor{lightcyan}{rgb}{0.75, 1.0, 1.0}
\definecolor{guppiegreen}{rgb}{0.0, 1.0, 0.5}
\definecolor{salmon}{rgb}{1.0, 0.55, 0.41}
\definecolor{strongsalmon}{rgb}{1.0,0.4,0.3}
\definecolor{lightsalmon}{rgb}{1.0,0.72,0.68}
\definecolor{samulightgrey}{rgb}{0.90, 0.90, 0.93}
\definecolor{trolleygrey}{rgb}{0.5, 0.5, 0.5}

\definecolor{salcyan0}{rgb}{0.65,0.9,0.8}
\definecolor{salcyan1}{rgb}{0.657070707071,0.896363636364,0.797575757576}
\definecolor{salcyan2}{rgb}{0.664141414141,0.892727272727,0.795151515152}
\definecolor{salcyan3}{rgb}{0.671212121212,0.889090909091,0.792727272727}
\definecolor{salcyan4}{rgb}{0.678282828283,0.885454545455,0.790303030303}
\definecolor{salcyan5}{rgb}{0.685353535354,0.881818181818,0.787878787879}
\definecolor{salcyan6}{rgb}{0.692424242424,0.878181818182,0.785454545455}
\definecolor{salcyan7}{rgb}{0.699494949495,0.874545454545,0.78303030303}
\definecolor{salcyan8}{rgb}{0.706565656566,0.870909090909,0.780606060606}
\definecolor{salcyan9}{rgb}{0.713636363636,0.867272727273,0.778181818182}
\definecolor{salcyan10}{rgb}{0.720707070707,0.863636363636,0.775757575758}
\definecolor{salcyan11}{rgb}{0.727777777778,0.86,0.773333333333}
\definecolor{salcyan12}{rgb}{0.734848484848,0.856363636364,0.770909090909}
\definecolor{salcyan13}{rgb}{0.741919191919,0.852727272727,0.768484848485}
\definecolor{salcyan14}{rgb}{0.74898989899,0.849090909091,0.766060606061}
\definecolor{salcyan15}{rgb}{0.756060606061,0.845454545455,0.763636363636}
\definecolor{salcyan16}{rgb}{0.763131313131,0.841818181818,0.761212121212}
\definecolor{salcyan17}{rgb}{0.770202020202,0.838181818182,0.758787878788}
\definecolor{salcyan18}{rgb}{0.777272727273,0.834545454545,0.756363636364}
\definecolor{salcyan19}{rgb}{0.784343434343,0.830909090909,0.753939393939}
\definecolor{salcyan20}{rgb}{0.791414141414,0.827272727273,0.751515151515}
\definecolor{salcyan21}{rgb}{0.798484848485,0.823636363636,0.749090909091}
\definecolor{salcyan22}{rgb}{0.805555555556,0.82,0.746666666667}
\definecolor{salcyan23}{rgb}{0.812626262626,0.816363636364,0.744242424242}
\definecolor{salcyan24}{rgb}{0.819696969697,0.812727272727,0.741818181818}
\definecolor{salcyan25}{rgb}{0.830303030303,0.807272727273,0.738181818182}
\definecolor{salcyan26}{rgb}{0.837373737374,0.803636363636,0.735757575758}
\definecolor{salcyan27}{rgb}{0.844444444444,0.8,0.733333333333}
\definecolor{salcyan28}{rgb}{0.851515151515,0.796363636364,0.730909090909}
\definecolor{salcyan29}{rgb}{0.858585858586,0.792727272727,0.728484848485}
\definecolor{salcyan30}{rgb}{0.865656565657,0.789090909091,0.726060606061}
\definecolor{salcyan31}{rgb}{0.872727272727,0.785454545455,0.723636363636}
\definecolor{salcyan32}{rgb}{0.879797979798,0.781818181818,0.721212121212}
\definecolor{salcyan33}{rgb}{0.886868686869,0.778181818182,0.718787878788}
\definecolor{salcyan34}{rgb}{0.893939393939,0.774545454545,0.716363636364}
\definecolor{salcyan35}{rgb}{0.90101010101,0.770909090909,0.713939393939}
\definecolor{salcyan36}{rgb}{0.908080808081,0.767272727273,0.711515151515}
\definecolor{salcyan37}{rgb}{0.915151515152,0.763636363636,0.709090909091}
\definecolor{salcyan38}{rgb}{0.922222222222,0.76,0.706666666667}
\definecolor{salcyan39}{rgb}{0.929292929293,0.756363636364,0.704242424242}
\definecolor{salcyan40}{rgb}{0.936363636364,0.752727272727,0.701818181818}
\definecolor{salcyan41}{rgb}{0.943434343434,0.749090909091,0.699393939394}
\definecolor{salcyan42}{rgb}{0.950505050505,0.745454545455,0.69696969697}
\definecolor{salcyan43}{rgb}{0.957575757576,0.741818181818,0.694545454545}
\definecolor{salcyan44}{rgb}{0.964646464646,0.738181818182,0.692121212121}
\definecolor{salcyan45}{rgb}{0.971717171717,0.734545454545,0.689696969697}
\definecolor{salcyan46}{rgb}{0.978787878788,0.730909090909,0.687272727273}
\definecolor{salcyan47}{rgb}{0.985858585859,0.727272727273,0.684848484848}
\definecolor{salcyan48}{rgb}{0.992929292929,0.723636363636,0.682424242424}
\definecolor{salcyan49}{rgb}{1.0,0.72,0.68}

\begin{table}[h]
\resizebox{\textwidth}{!}{%
\newcolumntype{?}{!{\vrule width2pt}}
\begin{tabular}{|c|c?c|c|c|c|c|c|c|c|}
\hline
\multicolumn{2}{|c?}{\multirow{4}{*}{D}} & \multicolumn{4}{c|}{link: add}   & \multicolumn{4}{c|}{link: del} \\ \cline{3-10}
\multicolumn{2}{|c?}{}  &  \multicolumn{2}{c|}{source: new} & \multicolumn{2}{c|}{source: old} & \multicolumn{2}{c|}{source: new} & \multicolumn{2}{c|}{source: old} \\ \cline{3-10}
\multicolumn{2}{|c?}{}    & target:  &   target:   &   target: &    target: &  target:  &   target: &  target:  & target:\\
\multicolumn{2}{|c?}{}     & new           &  old        &     new     &    old       &  new   &   old        &  new        &   old  \\ \thickhline
\parbox[t]{2mm}{\multirow{4}{*}{\rotatebox[origin=c]{90}{source}}}
& child.	& \cellcolor{lightsalmon} s+	& \cellcolor{lightsalmon} s+	& \cellcolor{lightsalmon} s+	& \cellcolor{lightsalmon} s+	& \cellcolor{samulightgrey}	& \cellcolor{samulightgrey}	& \cellcolor{samulightgrey}	& \cellcolor{lightsalmon} s+	 \\ \cline{2-10}
& par.	& \color{trolleygrey} i.s.	& \color{trolleygrey} i.s.	& \cellcolor{greyishcyan} s--	& \color{trolleygrey} i.s.	& \cellcolor{samulightgrey}	& \cellcolor{samulightgrey}	& \cellcolor{samulightgrey}	& \color{trolleygrey} i.s.	 \\ \cline{2-10}
& desc.	& \cellcolor{greyishcyan} p--	& \cellcolor{lightsalmon} s+	& \cellcolor{lightsalmon} s+	& \cellcolor{lightsalmon} s+	& \cellcolor{samulightgrey}	& \cellcolor{samulightgrey}	& \cellcolor{samulightgrey}	& \cellcolor{lightsalmon} p+	 \\ \cline{2-10}
& anc.	& s0	& \cellcolor{greyishcyan} s--	& \cellcolor{greyishcyan} s--	& \cellcolor{greyishcyan} s--	& \cellcolor{samulightgrey}	& \cellcolor{samulightgrey}	& \cellcolor{samulightgrey}	&  s0	 \\ \hline
\parbox[t]{2mm}{\multirow{4}{*}{\rotatebox[origin=c]{90}{target}}}
& child.	& \cellcolor{greyishcyan} s--	& \color{trolleygrey} i.s.	& \cellcolor{greyishcyan} s--	& s0	& \cellcolor{samulightgrey}	& \cellcolor{samulightgrey}	& \cellcolor{samulightgrey}	& s0	 \\ \cline{2-10}
& par.	& \color{trolleygrey} i.s.	& \color{trolleygrey} i.s.	& \cellcolor{lightsalmon} s+	& s0	& \cellcolor{samulightgrey}	& \cellcolor{samulightgrey}	& \cellcolor{samulightgrey}	& \cellcolor{lightsalmon} s+	 \\ \cline{2-10}
& desc.	& \cellcolor{greyishcyan} s--	& \color{trolleygrey} i.s.	& \cellcolor{greyishcyan} s--	& s0	& \cellcolor{samulightgrey}	& \cellcolor{samulightgrey}	& \cellcolor{samulightgrey}	& s0	 \\ \cline{2-10}
& anc.	& \cellcolor{lightsalmon} s+	& s0	& \cellcolor{lightsalmon} s+	& s0	& \cellcolor{samulightgrey}	& \cellcolor{samulightgrey}	& \cellcolor{samulightgrey}	& \cellcolor{lightsalmon} p+	 \\ \hline
\end{tabular}
}
\caption{ {\bf Summary of the results for hierarchy D}. The columns of the table correspond to the studied different link types, and the rows indicate the studied node property on either the source (top 4 rows) or the target (bottom 4 rows). The 3$^{\rm rd}$, 4$^{\rm th}$ and 5$^{\rm th}$ columns correspond to impossible link types, therefore, are left empty. The entries in the cells correspond to the following abbreviations: 's+', 's0' and 's-' for strong indication of preference, no preference and anti-preference, 'p+' and 'p-' for indication of preference or anti-preference with a peak, and 'i.s' for insufficient statistics.}
\label{table:hierarchy_D}
\end{table}

Summary tables analogous to Table \ref{table:hierarchy_D}. for the other hierarchies are listed in the Supporting Information. In order to be able to draw conclusions on the general features of the evolution of the studied hierarchies, we also provide an aggregated table with the same cell structure, in which the contribution from the individual tables were averaged in a simple manner, as shown in Table \ref{table:aggregated}. According to that we can make the following observations about the presence of preference or anti-preference with respect to the different node properties during the growth and restructuring of the studied hierarchies:
\newpage
\begin{itemize}
    \item We can see strong signs of preference with respect to the number of children of the source node for both the addition of new links pointing from old nodes to new ones, and for the deletion of already existing links between old nodes.
    \smallskip
    \item These link change events together with the addition of new links between already existing links clearly show preference with respect to the total number of descendants of the source node as well. 
	\smallskip    
    \item All possible link change types show anti-preference with respect to the total number of ancestors of the source node. This effect is strong in case of addition of new links with an old source node, and for adding new links pointing from new nodes to old ones, whereas can be considered somewhat less pronounced for new links between two new nodes, and relatively weak for link deletions. 
       \smallskip
    \item We can see both preference, neutral behaviour and anti-preference with respect to the total number of ancestors of the target node: the addition of new links pointing to new nodes and link deletions seem to display a weak preference, the addition of new links between old nodes displays neutral behaviour, whereas in case of the addition of new links pointing from new nodes to old ones, we can observe a weak anti-preference.
	\smallskip    
    \item The attachment/detachment processes seem to be more influenced by the properties of the source node of the changing links, compared to the influence of the properties of the target nodes. This is supported by the fact that the top 4 row in Table.\ref{table:aggregated}. contains much higher number of cells with values (other than 'i.s.'), and the magnitude of these is larger on average compared to cells in the bottom 4 rows.
   \smallskip    


\end{itemize}
An important further point to note is that the different hierarchies showed consistency in the sense that both preference and anti-preference was never observed simultaneously when comparing the same cells across the different summary tables. 

\definecolor{salcyan0}{rgb}{0.65,0.9,0.8}
\definecolor{salcyan1}{rgb}{0.657070707071,0.896363636364,0.797575757576}
\definecolor{salcyan2}{rgb}{0.664141414141,0.892727272727,0.795151515152}
\definecolor{salcyan3}{rgb}{0.671212121212,0.889090909091,0.792727272727}
\definecolor{salcyan4}{rgb}{0.678282828283,0.885454545455,0.790303030303}
\definecolor{salcyan5}{rgb}{0.685353535354,0.881818181818,0.787878787879}
\definecolor{salcyan6}{rgb}{0.692424242424,0.878181818182,0.785454545455}
\definecolor{salcyan7}{rgb}{0.699494949495,0.874545454545,0.78303030303}
\definecolor{salcyan8}{rgb}{0.706565656566,0.870909090909,0.780606060606}
\definecolor{salcyan9}{rgb}{0.713636363636,0.867272727273,0.778181818182}
\definecolor{salcyan10}{rgb}{0.720707070707,0.863636363636,0.775757575758}
\definecolor{salcyan11}{rgb}{0.727777777778,0.86,0.773333333333}
\definecolor{salcyan12}{rgb}{0.734848484848,0.856363636364,0.770909090909}
\definecolor{salcyan13}{rgb}{0.741919191919,0.852727272727,0.768484848485}
\definecolor{salcyan14}{rgb}{0.74898989899,0.849090909091,0.766060606061}
\definecolor{salcyan15}{rgb}{0.756060606061,0.845454545455,0.763636363636}
\definecolor{salcyan16}{rgb}{0.763131313131,0.841818181818,0.761212121212}
\definecolor{salcyan17}{rgb}{0.770202020202,0.838181818182,0.758787878788}
\definecolor{salcyan18}{rgb}{0.777272727273,0.834545454545,0.756363636364}
\definecolor{salcyan19}{rgb}{0.784343434343,0.830909090909,0.753939393939}
\definecolor{salcyan20}{rgb}{0.791414141414,0.827272727273,0.751515151515}
\definecolor{salcyan21}{rgb}{0.798484848485,0.823636363636,0.749090909091}
\definecolor{salcyan22}{rgb}{0.805555555556,0.82,0.746666666667}
\definecolor{salcyan23}{rgb}{0.812626262626,0.816363636364,0.744242424242}
\definecolor{salcyan24}{rgb}{0.819696969697,0.812727272727,0.741818181818}
\definecolor{salcyan25}{rgb}{0.830303030303,0.807272727273,0.738181818182}
\definecolor{salcyan26}{rgb}{0.837373737374,0.803636363636,0.735757575758}
\definecolor{salcyan27}{rgb}{0.844444444444,0.8,0.733333333333}
\definecolor{salcyan28}{rgb}{0.851515151515,0.796363636364,0.730909090909}
\definecolor{salcyan29}{rgb}{0.858585858586,0.792727272727,0.728484848485}
\definecolor{salcyan30}{rgb}{0.865656565657,0.789090909091,0.726060606061}
\definecolor{salcyan31}{rgb}{0.872727272727,0.785454545455,0.723636363636}
\definecolor{salcyan32}{rgb}{0.879797979798,0.781818181818,0.721212121212}
\definecolor{salcyan33}{rgb}{0.886868686869,0.778181818182,0.718787878788}
\definecolor{salcyan34}{rgb}{0.893939393939,0.774545454545,0.716363636364}
\definecolor{salcyan35}{rgb}{0.90101010101,0.770909090909,0.713939393939}
\definecolor{salcyan36}{rgb}{0.908080808081,0.767272727273,0.711515151515}
\definecolor{salcyan37}{rgb}{0.915151515152,0.763636363636,0.709090909091}
\definecolor{salcyan38}{rgb}{0.922222222222,0.76,0.706666666667}
\definecolor{salcyan39}{rgb}{0.929292929293,0.756363636364,0.704242424242}
\definecolor{salcyan40}{rgb}{0.936363636364,0.752727272727,0.701818181818}
\definecolor{salcyan41}{rgb}{0.943434343434,0.749090909091,0.699393939394}
\definecolor{salcyan42}{rgb}{0.950505050505,0.745454545455,0.69696969697}
\definecolor{salcyan43}{rgb}{0.957575757576,0.741818181818,0.694545454545}
\definecolor{salcyan44}{rgb}{0.964646464646,0.738181818182,0.692121212121}
\definecolor{salcyan45}{rgb}{0.971717171717,0.734545454545,0.689696969697}
\definecolor{salcyan46}{rgb}{0.978787878788,0.730909090909,0.687272727273}
\definecolor{salcyan47}{rgb}{0.985858585859,0.727272727273,0.684848484848}
\definecolor{salcyan48}{rgb}{0.992929292929,0.723636363636,0.682424242424}
\definecolor{salcyan49}{rgb}{1.0,0.72,0.68}

\begin{table}[ht!]
\resizebox{1.0\textwidth}{!}{%
\newcolumntype{?}{!{\vrule width2pt}}
\begin{tabular}{c!{\vrule width -0.4pt}c!{\vrule width -0.4pt}c!{\vrule width -0.4pt}c!{\vrule width -0.4pt}c!{\vrule width -0.4pt}c!{\vrule width -0.4pt}c!{\vrule width -0.4pt}c!{\vrule width -0.4pt}c!{\vrule width -0.4pt}c!{\vrule width -0.4pt}c!{\vrule width -0.4pt}c!{\vrule width -0.4pt}c!{\vrule width -0.4pt}c!{\vrule width -0.4pt}c!{\vrule width -0.4pt}c!{\vrule width -0.4pt}c!{\vrule width -0.4pt}c!{\vrule width -0.4pt}c!{\vrule width -0.4pt}c!{\vrule width -0.4pt}c!{\vrule width -0.4pt}c!{\vrule width -0.4pt}c!{\vrule width -0.4pt}c!{\vrule width -0.4pt}c!{\vrule width -0.4pt}c!{\vrule width -0.4pt}c!{\vrule width -0.4pt}c!{\vrule width -0.4pt}c!{\vrule width -0.4pt}c!{\vrule width -0.4pt}c!{\vrule width -0.4pt}c!{\vrule width -0.4pt}c!{\vrule width -0.4pt}c!{\vrule width -0.4pt}c!{\vrule width -0.4pt}c!{\vrule width -0.4pt}c!{\vrule width -0.4pt}c!{\vrule width -0.4pt}c!{\vrule width -0.4pt}c!{\vrule width -0.4pt}c!{\vrule width -0.4pt}c!{\vrule width -0.4pt}c!{\vrule width -0.4pt}c!{\vrule width -0.4pt}c!{\vrule width -0.4pt}c!{\vrule width -0.4pt}c!{\vrule width -0.4pt}c!{\vrule width -0.4pt}c!{\vrule width -0.4pt}c !{\vrule width -0.4pt}}  
\cellcolor{salcyan0} &	\cellcolor{salcyan1} &	\cellcolor{salcyan2} &	\cellcolor{salcyan3} &	\cellcolor{salcyan4} &	\cellcolor{salcyan5} &	\cellcolor{salcyan6} &	\cellcolor{salcyan7} &	\cellcolor{salcyan8} &	\cellcolor{salcyan9} &	\cellcolor{salcyan10} &	\cellcolor{salcyan11} &	\cellcolor{salcyan12} &	\cellcolor{salcyan13} &	\cellcolor{salcyan14} &	\cellcolor{salcyan15} &	\cellcolor{salcyan16} &	\cellcolor{salcyan17} &	\cellcolor{salcyan18} &	\cellcolor{salcyan19} &	\cellcolor{salcyan20} &	\cellcolor{salcyan21} &	\cellcolor{salcyan22} &	\cellcolor{salcyan23} &	\cellcolor{salcyan24} &	\cellcolor{salcyan25} &	\cellcolor{salcyan26} &	\cellcolor{salcyan27} &	\cellcolor{salcyan28} &	\cellcolor{salcyan29} &	\cellcolor{salcyan30} &	\cellcolor{salcyan31} &	\cellcolor{salcyan32} &	\cellcolor{salcyan33} &	\cellcolor{salcyan34} &	\cellcolor{salcyan35} &	\cellcolor{salcyan36} &	\cellcolor{salcyan37} &	\cellcolor{salcyan38} &	\cellcolor{salcyan39} &	\cellcolor{salcyan40} &	\cellcolor{salcyan41} &	\cellcolor{salcyan42} &	\cellcolor{salcyan43} &	\cellcolor{salcyan44} &	\cellcolor{salcyan45} &	\cellcolor{salcyan46} &	\cellcolor{salcyan47} &	\cellcolor{salcyan48} &	\cellcolor{salcyan49}  \\ 
 \cellcolor{white}\huge $\boldsymbol{s-}$ &  &	  &	  &	  &	  &	  &	  &	  &	  &	  &	  &	  &	  &	  &	  &	  &	  &	  &	  &	  &	  &	  &	  &	  &	\huge \textbf{neutral}  &	  &	  &	  &	  &	  &	  &	  &	  &	  &	  &	  &	  &	  &	  &	  &	  &	  &	  &	  &	  &	  &	  &	  &	\cellcolor{white}\huge $\boldsymbol{s+}$\\
 \cellcolor{white}  &	  &	  &	  &	  &	  &	  &	  &	  &	  &	  &	  &	  &	  &	  &	  &	  &	  &	  &	  &	  &	  &	  &	  &	  &	  &	  &	  &	  &	  &	  &	  &	  &	  &	  &	  &	  &	  &	  &	  &	  &	  &	  &	  &	  &	  &	  &	  &	  &   
\\
 \cellcolor{white}  &	  &	  &	  &	  &	  &	  &	  &	  &	  &	  &	  &	  &	  &	  &	  &	  &	  &	  &	  &	  &	  &	  &	  &	  &	  &	  &	  &	  &	  &	  &	  &	  &	  &	  &	  &	  &	  &	  &	  &	  &	  &	  &	  &	  &	  &	  &	  &	  &   
\end{tabular}
}
\resizebox{\textwidth}{!}{%
\newcolumntype{?}{!{\vrule width2pt}}
\begin{tabular}{|c|c?c|c|c|c|c|c|c|c|}
\hline
\multicolumn{2}{|c?}{\multirow{4}{*}{$\Sigma$}} & \multicolumn{4}{c|}{link: add}   & \multicolumn{4}{c|}{link: del} \\ \cline{3-10}
\multicolumn{2}{|c?}{}  &  \multicolumn{2}{c|}{source: new} & \multicolumn{2}{c|}{source: old} & \multicolumn{2}{c|}{source: new} & \multicolumn{2}{c|}{source: old} \\ \cline{3-10}
\multicolumn{2}{|c?}{}    & target:  &   target:   &   target: &    target: &  target:  &   target: &  target:  & target:\\
\multicolumn{2}{|c?}{}     & new           &  old        &     new     &    old       &  new   &   old        &  new        &   old  \\ \thickhline
\parbox[t]{2mm}{\multirow{4}{*}{\rotatebox[origin=c]{90}{source}}}
& child.	& \color{trolleygrey} i.s.	& \color{trolleygrey} i.s.	& \cellcolor{salcyan49}1.0,	& \color{trolleygrey} i.s.	& \cellcolor{samulightgrey}	& \cellcolor{samulightgrey}	& \cellcolor{samulightgrey}	& \cellcolor{salcyan41}0.67,	 \\ \cline{2-10}
& par.	& \color{trolleygrey} i.s.	& \color{trolleygrey} i.s.	& \color{trolleygrey} i.s.	& \color{trolleygrey} i.s.	& \cellcolor{samulightgrey}	& \cellcolor{samulightgrey}	& \cellcolor{samulightgrey}	& \color{trolleygrey} i.s.	 \\ \cline{2-10}
& desc.	& \color{trolleygrey} i.s.	& \color{trolleygrey} i.s.	& \cellcolor{salcyan46}0.89,	& \cellcolor{salcyan45}0.83,	& \cellcolor{samulightgrey}	& \cellcolor{samulightgrey}	& \cellcolor{samulightgrey}	& \cellcolor{salcyan39}0.61,	 \\ \cline{2-10}
& asc.	& \cellcolor{salcyan14}-0.44,	& \cellcolor{salcyan6}-0.75,	& \cellcolor{salcyan2}-0.90,	& \cellcolor{salcyan5}-0.78,	& \cellcolor{samulightgrey}	& \cellcolor{samulightgrey}	& \cellcolor{samulightgrey}	& \cellcolor{salcyan18}-0.28,	 \\ \hline
\parbox[t]{2mm}{\multirow{4}{*}{\rotatebox[origin=c]{90}{target}}}
& child.	& \color{trolleygrey} i.s.	& \color{trolleygrey} i.s.	& \color{trolleygrey} i.s.	& \color{trolleygrey} i.s.	& \cellcolor{samulightgrey}	& \cellcolor{samulightgrey}	& \cellcolor{samulightgrey}	& \color{trolleygrey} i.s.	 \\ \cline{2-10}
& par.	& \color{trolleygrey} i.s.	& \color{trolleygrey} i.s.	& \color{trolleygrey} i.s.	& \color{trolleygrey} i.s.	& \cellcolor{samulightgrey}	& \cellcolor{samulightgrey}	& \cellcolor{samulightgrey}	& \color{trolleygrey} i.s.	 \\ \cline{2-10}
& desc.	& \color{trolleygrey} i.s.	& \color{trolleygrey} i.s.	& \color{trolleygrey} i.s.	& \color{trolleygrey} i.s.	& \cellcolor{samulightgrey}	& \cellcolor{samulightgrey}	& \cellcolor{samulightgrey}	& \color{trolleygrey} i.s.	 \\ \cline{2-10}
& asc.	& \cellcolor{salcyan30}0.24,	& \cellcolor{salcyan22}-0.11,	& \cellcolor{salcyan31}0.29,	& \cellcolor{salcyan24}0.0,	& \cellcolor{samulightgrey}	& \cellcolor{samulightgrey}	& \cellcolor{samulightgrey}	& \cellcolor{salcyan31}0.29,	 \\ \hline
\end{tabular}
}
\caption{ {\bf Aggregated summary results}.
Based on table \ref{table:hierarchy_D}. and tables S8-S14. (given in the Supporting Information), the contribution to a given cell is counted according to 's+'=1, 'w+'='p+'=0.5, 's0=0', 'w--'='p--'=-0.5, 's--'=-1, and the obtained sum is divided by the number of tables contributing to the given cell. Aggregated cells become 'i.s' if more than 3 out of the 7 tables has 'i.s.' as well.}
\label{table:aggregated}
\end{table}

\newpage

\section*{Discussion}
We studied the change mechanisms of time evolving hierarchies between the PubMed MeSH terms using statistical methods. Although previous research has already shown interesting results regarding the growth of these networks \cite{Tsatsaronis,Leengheer,McCray,Tsatsaronis_2}, an important conclusion we can make based on our analysis is that deletion events and rewiring between already existing parts of the system are equally important in shaping the form of these hierarchies. This is supported by Tables S1-S7 in the Supporting Information, according to which the number of deleted links together with the number of new links between already existing nodes under one time step is usually of the same magnitude as the number of new links connected to newly appearing nodes. 

The main focus of our studies was on measuring preference during attachment and detachment events with respect to four different node properties characterising the hierarchy members. By setting up a general framework for this sort of analysis we could show that the likelihood for nodes to take part in restructuring events can be effected by their properties under quite a number of different circumstances. We found that when new links appear pointing from already existing nodes to newly appearing ones, the nodes with larger number of children (larger out degree) are chosen as source nodes for this type of links with significantly larger probabilities compared to uniform random choice. This effect is analogous to the preferential attachment rule of the Barabási--Albert network model \cite{BA_model}, which was also observed empirically in different growing network systems \cite{Newman-mer,Tamas-mer,Barab-mer,pref-coms}. However, in our case a larger number of children also increases the likelihood of loosing an out link (corresponding to a link deletion event). 

Another property for which we observed similar behaviour is the total number of descendants, where in addition to the above two effects we could also detect preference during the addition of a new links pointing to other already existing nodes. In parallel, we observed anti-preference with respect to the number of ancestors of the source node for all possible link change types. Interestingly, if the node acts as the target of the changing link, we can observe both preference and anti-preference with respect to the number of ancestors for the different link change types. Since the number of descendants and the number of ancestors are defined only in case of hierarchies, the related results have no previously observed analogy in general time dependent networks. 

Finally, we note that according to Table \ref{table:hierarchy_D}. and Tables S8-S14 in the Supporting Information one can observe a mild variance across the different hierarchies in terms of whether a given link change type displays some sort of preference  with respect to a given property, or we see a neutral behaviour (or insufficient statistics) instead. Nevertheless, the results across the different hierarchies are consistent in the sense that we cannot observe both preference in case of one hierarchy, and anti-preference in case of another hierarchy for the same link change type and node property. This consistency is encouraging from the point of view of further research focusing on building network models for time evolving hierarchies. In addition, we note that although the empirical studies in this work are restricted to the networks between MesH terms, it is quite plausible that a part of these features are more universal and occur in time evolution of networks with a hierarchical structure in general.

\section*{Conclusion}

In summary, our findings show that the growth and rewiring of the examined hierarchies are governed by non-trivial preference in the attachment mechanisms of the links. According to our results, the attachment is non-uniform with respect to multiple different topological and hierarchical node properties, and among the different possible link change scenarios we could observe both preferential and anti-preferential attachments, depending on the given node property of interest. These facts indicate that time evolution of these systems is far more complex compared to simple preferential attachment models, providing very interesting future challenges for modelling and further statistical analysis.


\section*{Supporting information}

\paragraph*{S1 File.}
\label{S1_File}
{\bf Supporting Information S1 (see below).}

\section*{Acknowledgments}
The research was partially supported by the European Union through projects 'RED-Alert' (grant no.: 740688-RED-Alert-H2020- SEC-2016-2017/H2020- SEC-2016-2017-1) and by the Hungarian National Research, Development and Innovation Office (grant no. K 128780).



%
%
%






\begin{thebibliography}{10}

\bibitem{Laci_revmod}
Albert R, Barab{\'a}si AL.
\newblock Statistical mechanics of complex networks.
\newblock Rev\ Mod\ Phys. 2002;74:47--97.

\bibitem{Dorog_book}
Mendes JFF, Dorogovtsev SN.
\newblock Evolution of Networks: From Biological Nets to the Internet and WWW.
\newblock Oxford: Oxford Univ.\ Press; 2003.

\bibitem{Laci_hier_scale}
Ravasz E, Somera AL, Mongru DA, Oltvai ZN, Barab\'asi AL.
\newblock Hierarchical Organization of Modularity in Metabolic Networks.
\newblock Science. 2002;297:1{55}1 -- 1{55}5.

\bibitem{Newman_hier}
Clauset A, Moore C, Newman MEJ.
\newblock Hierarchical structure and the prediction of missing links in
  networks.
\newblock Nature. 2008;453:98--101.

\bibitem{Pumain_book}
Pumain D.
\newblock Hierarchy in Natural and Social Sciences. vol.~3 of Methodos Series.
\newblock Dodrecht, The Netherlands: Springer Netherlands; 2006.

\bibitem{Sole_chaos_hier}
Corominas-Murtra B, Rodr{\'\i}guez-Caso C, Go{\~n}i J, Sol{\'e} R.
\newblock Measuring the hierarchy of feedforward networks.
\newblock Chaos. 2011;21:01{61}08.

\bibitem{Anna_and_Tamas_book}
Zafeiris A, Vicsek T.
\newblock Why We Live in Hierarchies? A Quantitative Treatise.
\newblock Berlin: Springer; 2018.

\bibitem{Sneppen_hier_measures}
Trusina A, Maslov S, Minnhagen P, Sneppen K.
\newblock Hierarchy measures in complex networks.
\newblock Phys\ Rev\ Lett. 2004;92:17{87}02.

\bibitem{Enys_hierarchy}
Mones E, Vicsek L, Vicsek T.
\newblock Hierarchy Measure for Complex Networks.
\newblock PLoS ONE. 2012;7:e33799.

\bibitem{Sole_hier_PNAS}
Corominas-Murtra B, Go{\~{n}}i J, Sol{\'e} RV, Rodríguez-Caso C.
\newblock On the origins of hierarchy in complex networks.
\newblock Proc\ Natl\ Acad\ Sci\ USA. 2013;110:13{31}6--–13{32}1.

\bibitem{RWH}
Cz{\'e}gel D, Palla G.
\newblock Random walk hierarchy: What is more hierarchical, a chain a tree or a
  star?
\newblock Scientific Reports. 2015;5:17994.

\bibitem{Gupte_hier_measure}
Gupte M, Shankar P, Li J, Muthukrishnan S, Iftode L.
\newblock Finding hierarchy in directed online social networks.
\newblock In: Proceedings of the 20th international conference on World wide
  web. ACM; 2011. p. 557--566.

\bibitem{Elisa_hier_measure}
Letizia E, Barucca P, Lillo F.
\newblock Resolution of ranking hierarchies in directed networks.
\newblock PLOS ONE. 2018;13(2):1--25.
\newblock doi:{10.1371/journal.pone.0191604}.

\bibitem{Zeng_Ecoli}
Ma HW, Buer J, Zeng AP.
\newblock Hierarchical sructure and modules in the Escherichia coli
  transcriptional regulatory network revealed by a new top-down approach.
\newblock BMC Bioinformatics. 2004;5:199.

\bibitem{Huber_crayfish}
Goessmann C, Hemelrijk C, Huber R.
\newblock The formation and maintenance of crayfish hierarchies: behavioral and
  self-structuring properties.
\newblock Behav\ Ecol\ Sociobiol. 2000;48:4{18}--–4{28}.

\bibitem{Tamas_pigeons}
Nagy M, Akos Z, Biro D, Vicsek T.
\newblock Hierarchical group dynamics in pigeon flocks.
\newblock Nature. 2010;464:8{90}--–8{93}.

\bibitem{Pigeon_context}
Nagy M, V{\'a}s{\'a}rhelyi G, Pettit B, Roberts-Mariani I, Vicsek T, Biro D.
\newblock Context-dependent hierarchies in pigeons.
\newblock Proc\ Natl\ Acad\ Sci\ USA. 2013;110:13{04}9--–13{05}4.

\bibitem{Ozogany}
Ozogány K, Vicsek T.
\newblock J Stat Phys. 2015;158:628.
\newblock doi:{https://doi.org/10.1007/s10955-014-1131-7}.

\bibitem{McCowan_macaque}
Fushing H, McAssey MP, Beisner B, McCowan B.
\newblock Ranking network of captive rhesus macaque society: A sophisticated
  corporative kingdom.
\newblock PLoS ONE. 2011;6:e17817.

\bibitem{Kaiser_neural}
Kaiser M, Hilgetag CC, K{\"o}tter R.
\newblock Hierarchy and dynamics of neural networks.
\newblock Front\ Neuroinform. 2010;4:112.

\bibitem{Palgrave}
Palla G, Tib{\'e}ly G, Mones E, Pollner P, Vicsek T.
\newblock Hierarchical networks of scientific journals.
\newblock Palgrave Communications. 2015;1:15016.

\bibitem{Guimera_hier_soc}
Guimer{\`a} R, Danon L, D{\'\i}az-Guilera A, Giralt F, Arenas A.
\newblock Self-similar community structure in a network of human interactions.
\newblock Phys\ Rev\ E. 2003;68:065103.

\bibitem{our_pref_coms}
Pollner P, Palla G, Vicsek T.
\newblock Preferential attachment of communities: The same principle, but a
  higher level.
\newblock Europhys\ Lett. 2006;73:4{78}--4{84}.

\bibitem{Sole_hier_soc}
Valverde S, Sol{\'e} RV.
\newblock Self-organization versus hierarchy in open-source social networks.
\newblock Phys\ Rev\ E. 2007;76:046118.

\bibitem{PicturAsk}
Tóth BJ, Palla G, Mones E, Havadi G, Páll N, Pollner P, et~al.
\newblock Emergence of Leader-Follower Hierarchy Among Players in an On-Line
  Experiment.
\newblock In: 2018 IEEE/ACM International Conference on Advances in Social
  Networks Analysis and Mining (ASONAM); 2018. p. 1184--1190.

\bibitem{Krugman_urban}
Krugman PR.
\newblock Confronting the mystery of urban hierarchy.
\newblock J\ Jpn\ Int\ Econ. 1996;10:3{99}--–{41}8.

\bibitem{Batty_urban}
Batty M, Longley P.
\newblock Fractal Cities: A Geometry of Form and Function.
\newblock San Diego: Academic; 1994.

\bibitem{news_portals}
Tib{\'e}ly G, Sousa-Rodrigues D, Pollner P, Palla G.
\newblock Comparing the Hierarchy of Keywords in On-Line News Portals.
\newblock PLoS ONE. 2016;11:e0165728.

\bibitem{Hirata_eco}
Hirata H, Ulanowicz R.
\newblock Information theoretical analysis of the aggregation and hierarchical
  structure of ecological networks.
\newblock J\ Theor\ Biol. 1985;116:3{21}–--{34}1.

\bibitem{Wickens_eco}
Wickens J, Ulanowicz R.
\newblock On quantifying hierarchical connections in ecology.
\newblock J\ Soc\ Biol\ Struct. 1988;11:3{69}--–{37}8.

\bibitem{Eldrege_book}
Eldredge N.
\newblock Unfinished Synthesis: Biological Hierarchies and Modern Evolutionary
  Thought.
\newblock New York: Oxford Univ.\ Press; 1985.

\bibitem{McShea_organism}
McShea DW.
\newblock The hierarchical structure of organisms.
\newblock Paleobiology. 2001;27:4{05}--–{42}3.

\bibitem{Mengistu_evolv_hier}
Mengistu H, Huizinga J, Mouret JB, Clune J.
\newblock The Evolutionary Origins of Hierarchy.
\newblock PLOS Computational Biology. 2016;12(6):1--23.
\newblock doi:{10.1371/journal.pcbi.1004829}.

\bibitem{Lambiotte_non_normal}
Asllani M, Lambiotte R, Carletti T.
\newblock Structure and dynamical behavior of non-normal networks.
\newblock Science Advances. 2018;4(12).
\newblock doi:{10.1126/sciadv.aau9403}.

\bibitem{Katchanov2017}
Katchanov YL, Markova YV.
\newblock The ``space of physics journals'': topological structure and the
  Journal Impact Factor.
\newblock Scientometrics. 2017;113(1):313--333.
\newblock doi:{10.1007/s11192-017-2471-2}.

\bibitem{science_mapping}
Chen C.
\newblock Science Mapping: A Systematic Review of the Literature.
\newblock Journal of Data and Information Science. 2017;2(2):1--40.
\newblock doi:{https://doi.org/10.1515/jdis-2017-0006}.

\bibitem{Chen_book_chap}
Chen C, Song M.
\newblock In: Science Mapping Tools and Applications. Springer, Cham; 2017.

\bibitem{graphene_paper}
Alvial-Palavicino C, Konrad K.
\newblock The rise of graphene expectations: Anticipatory practices in emergent
  nanotechnologies.
\newblock Futures. 2018;doi:{https://doi.org/10.1016/j.futures.2018.10.008}.

\bibitem{sustainability_paper}
Shan W, Wang J.
\newblock Mapping the Landscape and Evolutions of Green Supply Chain
  Management.
\newblock Sustainability. 2018;10(3).
\newblock doi:{10.3390/su10030597}.

\bibitem{Anna_Tamas_Nature_Communications}
Zafeiris A, Vicsek T.
\newblock Group performance is maximized by hierarchical competence
  distribution.
\newblock Nature Communications. 2013;4:2484.
\newblock doi:{https://doi.org/10.1038/ncomms3484}.

\bibitem{Zamani_2017}
Zamani M, Vicsek T.
\newblock Glassy nature of hierarchical organizations.
\newblock Scientific Reports. 2017;7:1382.
\newblock doi:{10.1038/s41598-017-01503-y}.

\bibitem{Zamani_2018}
Zamani M, Camargo-Forero L, Vicsek T.
\newblock Stability of glassy hierarchical networks.
\newblock New Journal of Physics. 2018;20(2):023025.
\newblock doi:{10.1088/1367-2630/aaa8ca}.

\bibitem{BA_model}
Barab{\'a}si AL, Albert R.
\newblock Emergence of scaling in random networks.
\newblock Science. 1999;286:509--512.

\bibitem{Tamas-mer}
Barab\'asi AL, Jeong H, N\'eda Z, Ravasz E, Schubert A, Vicsek T.
\newblock Evolution of the social network of scientific collaborations.
\newblock Physica A. 2002;311:590--614.

\bibitem{Barab-mer}
Jeong H, N{\'e}da Z, Barab{\'a}si AL.
\newblock Measuring preferential attachment in evolving networks.
\newblock EPL. 2003;61:567.

\bibitem{Newman-mer}
Newman MEJ.
\newblock Clustering and preferential attachment in growing networks.
\newblock Phys\ Rev\ E. 2001;64:02{51}02(R).

\bibitem{pref-coms}
Pollner P, Palla G, Vicsek T.
\newblock Preferential attachment of communities: The same principle, but a
  higher level.
\newblock EPL. 2006;73:478.

\bibitem{group_evolv_nature}
Palla G, Barab\'asi AL, Vicsek T.
\newblock Quantifying social group evolution.
\newblock Nature. 2007;446:664--667.

\bibitem{Tsatsaronis}
Tsatsaronis G, Macari N, Torge S, Dietze H, Schroeder M.
\newblock A Maximum-Entropy approach for accurate document annotation in the
  biomedical domain.
\newblock Journal of Biomedical Semantics. 2012;3:S2.

\bibitem{Leengheer}
de~Leenheer P.
\newblock 5.
\newblock In: Hepp M, de~Leenheer P, Moor AD, Sure Y, editors. Ontology
  evolution. US: Springer; 2008. p. 131--176.

\bibitem{McCray}
McCray AT, Lee K.
\newblock In: K{\"u}ppers BO, Hahn U, Artmann S, editors. Taxonomic Change as a
  Reflection of Progress in a Scientific Discipline. Berlin, Heidelberg:
  Springer Berlin Heidelberg; 2013. p. 189--208.

\bibitem{Tsatsaronis_2}
Tsatsaronis G, Varlamis I, Kanhabua N, N{\o}rv{\aa}g K.
\newblock In: Gelbukh A, editor. Temporal Classifiers for Predicting the
  Expansion of Medical Subject Headings. Berlin, Heidelberg: Springer Berlin
  Heidelberg; 2013. p. 98--113.

\bibitem{raw_MeSH_data}
In this study we use publicly available data from the website of PubMed;.
\newblock Available from: \url{https://www.nlm.nih.gov/mesh/filelist.html}.

\end{thebibliography}

\renewcommand{\thefigure}{\arabic{figure}}
\renewcommand{\thetable}{\arabic{table}}
\renewcommand{\theequation}{S\arabic{equation}}
\renewcommand{\thesection}{S\arabic{section}}

\renewcommand{\thetable}{\Alph{table}}
\renewcommand{\thefigure}{\Alph{figure}}


\part*{Supporting information}

\section{Basic properties of the MeSH hierarchies}

Owing to the yearly updates, the MeSH hierarchies evolve in time, displaying great number of structural changes that affect their topology both on the level of nodes and on the level of links. We studied the number of different annual change event types for each hierarchy with the corresponding results given in \ref{tab:annualchangeA}-\ref{tab:annualchangeN} Tables.


\begin{table}[ht!]
\resizebox{\textwidth}{!}{%
\newcolumntype{?}{!{\vrule width2pt}}
\begin{tabular}{|c|c|c|c|c|c|c|c|c|}
\cline{1-5} \cline{6-9} 
\multirow{3}{*}{year} & \multirow{3}{*}{size} & num. of & num. of & num. of & \multicolumn{4}{c|}{num. of}  \\
&  & del. nodes & add. nodes & del. links & \multicolumn{4}{c|}{add. links} \\ \cline{6-9}
 &  &  &  & & $o\to o$ & $o\to n$ & $n\to o$ & $n\to n$\\ \hline
 2002 & 1350 & 0 & 40 & 29 & 11 & 38 & 25 & 7 \\ \hline
 2003 & 1390 & 1 & 34 & 49 & 27 & 39 & 38 & 4 \\ \hline
 2004 & 1423 & 1 & 11 & 5 & 6 & 17 & 1 & 1 \\ \hline
 2005 & 1433 & 3 & 38 & 39 & 12 & 41 & 20 & 8 \\ \hline
 2006 & 1468 & 5 & 29 & 10 & 6 & 28 & 2 & 9 \\ \hline
 2007 & 1492 & 1 & 38 & 35 & 36 & 34 & 5 & 11 \\ \hline
 2008 & 1529 & 4 & 33 & 36 & 14 & 38 & 30 & 4 \\ \hline
 2009 & 1558 & 0 & 58 & 0 & 4 & 24 & 10 & 47 \\ \hline
 2010 & 1616 & 1 & 32 & 3 & 0 & 40 & 3 & 2 \\ \hline
 2011 & 1647 & 1 & 30 & 5 & 10 & 32 & 6 & 1 \\ \hline
 2012 & 1676 & 1 & 9 & 1 & 1 & 10 & 0 & 0 \\ \hline
 2013 & 1684 & 1 & 20 & 1 & 0 & 12 & 1 & 10 \\ \hline
 2014 & 1703 & 0 & 61 & 23 & 9 & 58 & 23 & 16 \\ \hline
 2015 & 1764 & 1 & 17 & 27 & 11 & 27 & 3 & 3 \\ \hline
 2016 & 1780 & 0 & 39 & 9 & 5 & 41 & 5 & 0 \\ \hline
 2017 & 1819 & 0 & 7 & 0 & 2 & 11 & 0 & 0 \\ \hline
\end{tabular}
}
\caption{\textbf{Number of different annual change event types in hierarchy A.} The 1$^{\rm st}$ column displays the year of observation with the corresponding network size given in the 2$^{\rm nd}$ column. The 3$^{\rm rd}$ and 4$^{\rm th}$ columns correspond to the number of deleted and added nodes within that particular year. The 5$^{\rm th}$ column displays the number of deleted links while the 6$^{\rm th}$, 7$^{\rm th}$, 8$^{\rm th}$, 9$^{\rm th}$ columns correspond to the number of link addition between old nodes, between old sources and new targets, between new sources and old targets, and between new nodes.}
\label{tab:annualchangeA}
\end{table}

\begin{table}[ht!]
\resizebox{\textwidth}{!}{%
\newcolumntype{?}{!{\vrule width2pt}}
\begin{tabular}{|c|c|c|c|c|c|c|c|c|}
\cline{1-5} \cline{6-9} 
\multirow{3}{*}{year} & \multirow{3}{*}{size} & num. of & num. of & num. of & \multicolumn{4}{c|}{num. of}  \\
&  & del. nodes & add. nodes & del. links & \multicolumn{4}{c|}{add. links} \\ \cline{6-9}
 &  &  &  & & $o\to o$ & $o\to n$ & $n\to o$ & $n\to n$\\ \hline
 2002 & 2252 & 9 & 722 & 85 & 37 & 622 & 46 & 104 \\ \hline
 2003 & 2965 & 19 & 243 & 156 & 53 & 252 & 61 & 38 \\ \hline
 2004 & 3189 & 9 & 171 & 48 & 22 & 157 & 27 & 17 \\ \hline
 2005 & 3351 & 12 & 96 & 39 & 5 & 60 & 22 & 37 \\ \hline
 2006 & 3435 & 15 & 62 & 51 & 18 & 62 & 12 & 5 \\ \hline
 2007 & 3482 & 0 & 16 & 3 & 5 & 16 & 1 & 1 \\ \hline
 2008 & 3498 & 1 & 57 & 7 & 4 & 63 & 2 & 7 \\ \hline
 2009 & 3554 & 40 & 57 & 81 & 3 & 26 & 31 & 42 \\ \hline
 2010 & 3571 & 2 & 62 & 19 & 9 & 57 & 8 & 12 \\ \hline
 2011 & 3631 & 0 & 25 & 7 & 0 & 29 & 7 & 2 \\ \hline
 2012 & 3656 & 3 & 4 & 5 & 2 & 4 & 0 & 0 \\ \hline
 2013 & 3657 & 0 & 13 & 3 & 0 & 8 & 2 & 5 \\ \hline
 2014 & 3670 & 1 & 14 & 2 & 0 & 13 & 0 & 1 \\ \hline
 2015 & 3683 & 4 & 27 & 76 & 26 & 24 & 37 & 4 \\ \hline
 2016 & 3706 & 3 & 68 & 54 & 7 & 60 & 57 & 11 \\ \hline
 2017 & 3771 & 5 & 49 & 180 & 78 & 36 & 98 & 13 \\ \hline
\end{tabular}
}
\caption{\textbf{Number of different annual change event types in hierarchy B.} The arrangement of the table is the same as in case of \ref{tab:annualchangeA} Table: 1$^{\rm st}$ column is the year, 2$^{\rm nd}$ column is the size, 3$^{\rm rd}$ column is for the deleted nodes,  4$^{\rm th}$ column is for the added nodes, 5$^{\rm th}$ column is for the deleted links, and the columns from 6$^{\rm th}$ to 9$^{\rm th}$  display the different types of added links.}
\label{tab:annualchangeB}
\end{table}


\begin{table}[ht!]
\resizebox{\textwidth}{!}{%
\newcolumntype{?}{!{\vrule width2pt}}
\begin{tabular}{|c|c|c|c|c|c|c|c|c|}
\cline{1-5} \cline{6-9} 
\multirow{3}{*}{year} & \multirow{3}{*}{size} & num. of & num. of & num. of & \multicolumn{4}{c|}{num. of}  \\
&  & del. nodes & add. nodes & del. links & \multicolumn{4}{c|}{add. links} \\ \cline{6-9}
 &  &  &  & & $o\to o$ & $o\to n$ & $n\to o$ & $n\to n$\\ \hline
 2002 & 3975 & 3 & 43 & 28 & 20 & 58 & 39 & 3 \\ \hline
 2003 & 4015 & 5 & 44 & 99 & 59 & 58 & 25 & 4 \\ \hline
 2004 & 4054 & 5 & 61 & 82 & 50 & 68 & 22 & 6 \\ \hline
 2005 & 4110 & 11 & 65 & 79 & 31 & 77 & 47 & 10 \\ \hline
 2006 & 4164 & 5 & 71 & 95 & 73 & 95 & 24 & 6 \\ \hline
 2007 & 4230 & 34 & 73 & 153 & 72 & 94 & 18 & 11 \\ \hline
 2008 & 4269 & 4 & 59 & 45 & 23 & 70 & 19 & 11 \\ \hline
 2009 & 4324 & 6 & 91 & 43 & 22 & 161 & 30 & 10 \\ \hline
 2010 & 4409 & 6 & 92 & 45 & 36 & 139 & 18 & 7 \\ \hline
 2011 & 4495 & 0 & 83 & 16 & 19 & 125 & 21 & 13 \\ \hline
 2012 & 4578 & 4 & 23 & 9 & 6 & 32 & 0 & 1 \\ \hline
 2013 & 4597 & 4 & 28 & 35 & 15 & 34 & 14 & 3 \\ \hline
 2014 & 4621 & 3 & 46 & 21 & 4 & 61 & 9 & 6 \\ \hline
 2015 & 4664 & 0 & 23 & 103 & 26 & 28 & 3 & 0 \\ \hline
 2016 & 4687 & 0 & 72 & 21 & 19 & 91 & 30 & 6 \\ \hline
 2017 & 4759 & 0 & 40 & 32 & 17 & 61 & 30 & 0 \\ \hline
\end{tabular}
}
\caption{\textbf{Number of different annual change event types in hierarchy C.} The arrangement of the table is the same as in case of \ref{tab:annualchangeA} Table: 1$^{\rm st}$ column is the year, 2$^{\rm nd}$ column is the size, 3$^{\rm rd}$ column is for the deleted nodes,  4$^{\rm th}$ column is for the added nodes, 5$^{\rm th}$ column is for the deleted links, and columns from 6$^{\rm th}$ to 9$^{\rm th}$  display the different types of added links.}
\label{tab:annualchangeC}
\end{table}

\begin{table}[ht!]
\resizebox{\textwidth}{!}{%
\newcolumntype{?}{!{\vrule width2pt}}
\begin{tabular}{|c|c|c|c|c|c|c|c|c|}
\cline{1-5} \cline{6-9} 
\multirow{3}{*}{year} & \multirow{3}{*}{size} & num. of & num. of & num. of & \multicolumn{4}{c|}{num. of}  \\
&  & del. nodes & add. nodes & del. links & \multicolumn{4}{c|}{add. links} \\ \cline{6-9}
 &  &  &  & & $o\to o$ & $o\to n$ & $n\to o$ & $n\to n$\\ \hline
 2002 & 6902 & 10 & 251 & 166 & 74 & 225 & 86 & 148 \\ \hline
 2003 & 7143 & 56 & 269 & 329 & 111 & 235 & 181 & 106 \\ \hline
 2004 & 7356 & 23 & 142 & 109 & 76 & 119 & 76 & 55 \\ \hline
 2005 & 7475 & 34 & 700 & 389 & 224 & 651 & 154 & 330 \\ \hline
 2006 & 8141 & 5 & 269 & 141 & 95 & 263 & 43 & 113 \\ \hline
 2007 & 8405 & 6 & 219 & 72 & 78 & 200 & 51 & 87 \\ \hline
 2008 & 8618 & 1 & 103 & 21 & 12 & 88 & 13 & 53 \\ \hline
 2009 & 8720 & 5 & 101 & 43 & 24 & 99 & 17 & 53 \\ \hline
 2010 & 8816 & 7 & 166 & 64 & 26 & 152 & 64 & 63 \\ \hline
 2011 & 8975 & 14 & 115 & 64 & 22 & 124 & 33 & 38 \\ \hline
 2012 & 9076 & 23 & 108 & 182 & 113 & 100 & 74 & 53 \\ \hline
 2013 & 9161 & 3 & 122 & 76 & 34 & 124 & 48 & 40 \\ \hline
 2014 & 9280 & 1 & 74 & 18 & 17 & 65 & 10 & 30 \\ \hline
 2015 & 9353 & 1 & 165 & 101 & 20 & 224 & 13 & 42 \\ \hline
 2016 & 9517 & 1 & 234 & 52 & 56 & 267 & 72 & 77 \\ \hline
 2017 & 9750 & 2 & 186 & 198 & 56 & 279 & 39 & 36 \\ \hline
\end{tabular}
}
\caption{\textbf{Number of different annual change event types in hierarchy D.} The arrangement of the table is the same as in case of \ref{tab:annualchangeA} Table: 1$^{\rm st}$ column is the year, 2$^{\rm nd}$ column is the size, 3$^{\rm rd}$ column is for the deleted nodes,  4$^{\rm th}$ column is for the added nodes, 5$^{\rm th}$ column is for the deleted links, and columns from 6$^{\rm th}$ to 9$^{\rm th}$ display the different types of added links.}
\label{tab:annualchangeD}
\end{table}

\begin{table}[ht!]
\resizebox{\textwidth}{!}{%
\newcolumntype{?}{!{\vrule width2pt}}
\begin{tabular}{|c|c|c|c|c|c|c|c|c|}
\cline{1-5} \cline{6-9} 
\multirow{3}{*}{year} & \multirow{3}{*}{size} & num. of & num. of & num. of & \multicolumn{4}{c|}{num. of}  \\
&  & del. nodes & add. nodes & del. links & \multicolumn{4}{c|}{add. links} \\ \cline{6-9}
 &  &  &  & & $o\to o$ & $o\to n$ & $n\to o$ & $n\to n$\\ \hline
 2002 & 2040 & 5 & 57 & 19 & 6 & 60 & 12 & 6 \\ \hline
 2003 & 2092 & 0 & 28 & 5 & 7 & 28 & 4 & 3 \\ \hline
 2004 & 2120 & 3 & 59 & 33 & 22 & 43 & 23 & 26 \\ \hline
 2005 & 2176 & 5 & 29 & 32 & 22 & 33 & 6 & 1 \\ \hline
 2006 & 2200 & 3 & 43 & 12 & 6 & 47 & 10 & 6 \\ \hline
 2007 & 2240 & 4 & 31 & 36 & 15 & 35 & 12 & 3 \\ \hline
 2008 & 2267 & 5 & 113 & 30 & 16 & 79 & 32 & 49 \\ \hline
 2009 & 2375 & 0 & 70 & 25 & 28 & 78 & 6 & 5 \\ \hline
 2010 & 2445 & 5 & 113 & 39 & 18 & 133 & 39 & 8 \\ \hline
 2011 & 2553 & 6 & 83 & 63 & 57 & 103 & 14 & 6 \\ \hline
 2012 & 2630 & 1 & 74 & 8 & 2 & 79 & 15 & 7 \\ \hline
 2013 & 2703 & 4 & 28 & 15 & 6 & 28 & 5 & 2 \\ \hline
 2014 & 2727 & 0 & 34 & 23 & 13 & 37 & 1 & 0 \\ \hline
 2015 & 2761 & 3 & 60 & 71 & 14 & 62 & 14 & 10 \\ \hline
 2016 & 2818 & 4 & 51 & 20 & 11 & 62 & 1 & 1 \\ \hline
 2017 & 2865 & 1 & 60 & 9 & 9 & 64 & 9 & 10 \\ \hline
\end{tabular}
}
\caption{\textbf{Number of different annual change event types in hierarchy E.} The arrangement of the table is the same as in case of \ref{tab:annualchangeA} Table: 1$^{\rm st}$ column is the year, 2$^{\rm nd}$ column is the size, 3$^{\rm rd}$ column is for the deleted nodes,  4$^{\rm th}$ column is for the added nodes, 5$^{\rm th}$ column is for the deleted links, and columns from 6$^{\rm th}$ to 9$^{\rm th}$ display the different types of added links.}
\label{tab:annualchangeE}
\end{table}

\begin{table}[ht!]
\resizebox{\textwidth}{!}{%
\newcolumntype{?}{!{\vrule width2pt}}
\begin{tabular}{|c|c|c|c|c|c|c|c|c|}
\cline{1-5} \cline{6-9} 
\multirow{3}{*}{year} & \multirow{3}{*}{size} & num. of & num. of & num. of & \multicolumn{4}{c|}{num. of}  \\
&  & del. nodes & add. nodes & del. links & \multicolumn{4}{c|}{add. links} \\ \cline{6-9}
 &  &  &  & & $o\to o$ & $o\to n$ & $n\to o$ & $n\to n$\\ \hline
 2002 & 1803 & 4 & 112 & 111 & 52 & 50 & 83 & 78 \\ \hline
 2003 & 1911 & 32 & 84 & 159 & 31 & 50 & 94 & 47 \\ \hline
 2004 & 1963 & 17 & 48 & 84 & 41 & 46 & 33 & 9 \\ \hline
 2005 & 1994 & 19 & 52 & 79 & 27 & 54 & 26 & 5 \\ \hline
 2006 & 2027 & 8 & 64 & 69 & 21 & 57 & 59 & 11 \\ \hline
 2007 & 2083 & 3 & 38 & 23 & 24 & 39 & 4 & 4 \\ \hline
 2008 & 2118 & 632 & 247 & 1059 & 123 & 65 & 345 & 229 \\ \hline
 2009 & 1733 & 4 & 37 & 29 & 17 & 38 & 16 & 4 \\ \hline
 2010 & 1766 & 4 & 75 & 13 & 11 & 89 & 4 & 16 \\ \hline
 2011 & 1837 & 6 & 91 & 26 & 16 & 108 & 14 & 16 \\ \hline
 2012 & 1922 & 0 & 28 & 4 & 0 & 31 & 2 & 2 \\ \hline
 2013 & 1950 & 0 & 28 & 19 & 4 & 34 & 12 & 1 \\ \hline
 2014 & 1978 & 1 & 48 & 10 & 7 & 53 & 3 & 16 \\ \hline
 2015 & 2025 & 3 & 170 & 54 & 12 & 45 & 22 & 145 \\ \hline
 2016 & 2192 & 44 & 68 & 483 & 410 & 67 & 14 & 17 \\ \hline
 2017 & 2216 & 0 & 43 & 10 & 6 & 45 & 11 & 3 \\ \hline
\end{tabular}
}
\caption{\textbf{Number of different annual change event types in hierarchy G.} The arrangement of the table is the same as in case of  \ref{tab:annualchangeA} Table: 1$^{\rm st}$ column is the year, 2$^{\rm nd}$ column is the size, 3$^{\rm rd}$ column is for the deleted nodes,  4$^{\rm th}$ column is for the added nodes, 5$^{\rm th}$ column is for the deleted links, and columns from 6$^{\rm th}$ to 9$^{\rm th}$ display the different types of added links.}
\label{tab:annualchangeG}
\end{table}

\begin{table}[ht!]
\resizebox{\textwidth}{!}{%
\newcolumntype{?}{!{\vrule width2pt}}
\begin{tabular}{|c|c|c|c|c|c|c|c|c|}
\cline{1-5} \cline{6-9} 
\multirow{3}{*}{year} & \multirow{3}{*}{size} & num. of & num. of & num. of & \multicolumn{4}{c|}{num. of}  \\
&  & del. nodes & add. nodes & del. links & \multicolumn{4}{c|}{add. links} \\ \cline{6-9}
 &  &  &  & & $o\to o$ & $o\to n$ & $n\to o$ & $n\to n$\\ \hline
 2002 & 1072 & 0 & 36 & 1 & 0 & 39 & 7 & 2 \\ \hline
 2003 & 1108 & 0 & 5 & 6 & 4 & 3 & 2 & 2 \\ \hline
 2004 & 1113 & 1 & 7 & 1 & 0 & 7 & 0 & 0 \\ \hline
 2005 & 1119 & 1 & 3 & 2 & 0 & 4 & 1 & 0 \\ \hline
 2006 & 1121 & 0 & 1 & 0 & 0 & 1 & 0 & 0 \\ \hline
 2007 & 1122 & 2 & 38 & 22 & 11 & 37 & 4 & 4 \\ \hline
 2008 & 1158 & 0 & 254 & 6 & 11 & 43 & 20 & 244 \\ \hline
 2009 & 1412 & 0 & 31 & 3 & 16 & 34 & 0 & 4 \\ \hline
 2010 & 1443 & 0 & 43 & 2 & 3 & 45 & 4 & 5 \\ \hline
 2011 & 1486 & 0 & 38 & 5 & 2 & 41 & 5 & 1 \\ \hline
 2012 & 1524 & 10 & 52 & 16 & 0 & 50 & 3 & 7 \\ \hline
 2013 & 1566 & 3 & 34 & 4 & 1 & 38 & 0 & 2 \\ \hline
 2014 & 1597 & 1 & 25 & 5 & 5 & 24 & 0 & 3 \\ \hline
 2015 & 1621 & 3 & 79 & 35 & 18 & 85 & 5 & 9 \\ \hline
 2016 & 1697 & 9 & 66 & 24 & 10 & 68 & 4 & 6 \\ \hline
 2017 & 1754 & 0 & 41 & 8 & 4 & 46 & 4 & 1 \\ \hline
\end{tabular}
}
\caption{\textbf{Number of different annual change event types in hierarchy~N.} The arrangement of the table is the same as in case of \ref{tab:annualchangeA} Table: 1$^{\rm st}$ column is the year, 2$^{\rm nd}$ column is the size, 3$^{\rm rd}$ column is for the deleted nodes,  4$^{\rm th}$ column is for the added nodes, 5$^{\rm th}$ column is for the deleted links, and columns from 6$^{\rm th}$ to 9$^{\rm th}$ display the different types of added links.}
\label{tab:annualchangeN}
\end{table}

\definecolor{cyan}{rgb}{0.0, 1.0, 1.0}
\definecolor{greyishcyan}{rgb}{0.65, 0.9, 0.8}
\definecolor{lightcyan}{rgb}{0.75, 1.0, 1.0}
\definecolor{guppiegreen}{rgb}{0.0, 1.0, 0.5}
\definecolor{salmon}{rgb}{1.0, 0.55, 0.41}
\definecolor{strongsalmon}{rgb}{1.0,0.4,0.3}
\definecolor{lightsalmon}{rgb}{1.0,0.72,0.68}
\definecolor{samulightgrey}{rgb}{0.90, 0.90, 0.93}
\definecolor{trolleygrey}{rgb}{0.5, 0.5, 0.5}

\newpage

\section{The overall pattern of the different preference types}

The observed different preference types among all possible link change scenarios are listed in \ref{tab:Apreftab}-\ref{tab:Npreftab} Tables, each of which corresponds to a different hierarchy.

\begin{table}[h]
\newcolumntype{?}{!{\vrule width2pt}}
\begin{tabular}{|c|c?c|c|c|c|c|c|c|c|}
\hline
\multicolumn{2}{|c?}{\multirow{4}{*}{A}} & \multicolumn{4}{c|}{link: add}   & \multicolumn{4}{c|}{link: del} \\ \cline{3-10}
\multicolumn{2}{|c?}{}  &  \multicolumn{2}{c|}{source: new} & \multicolumn{2}{c|}{source: old} & \multicolumn{2}{c|}{source: new} & \multicolumn{2}{c|}{source: old} \\ \cline{3-10}
\multicolumn{2}{|c?}{}    & target:  &   target:   &   target: &    target: &  target:  &   target: &  target:  & target:\\
\multicolumn{2}{|c?}{}     & new           &  old        &     new     &    old       &  new   &   old        &  new        &   old  \\ \thickhline
\parbox[t]{2mm}{\multirow{4}{*}{\rotatebox[origin=c]{90}{source}}}
& child.	& \color{trolleygrey} i.s.	& \color{trolleygrey} i.s.	& \cellcolor{lightsalmon} s+	& \color{trolleygrey} i.s.	& \cellcolor{samulightgrey}	& \cellcolor{samulightgrey}	& \cellcolor{samulightgrey}	& \cellcolor{lightsalmon}w+	 \\ \cline{2-10}
& par.	& \color{trolleygrey} i.s.	& \color{trolleygrey} i.s.	& \color{trolleygrey} i.s.	& \color{trolleygrey} i.s.	& \cellcolor{samulightgrey}	& \cellcolor{samulightgrey}	& \cellcolor{samulightgrey}	& \color{trolleygrey} i.s.	 \\ \cline{2-10}
& desc.	& \color{trolleygrey} i.s.	& \cellcolor{lightsalmon}w+	& \cellcolor{lightsalmon} p+	& \color{trolleygrey} i.s.	& \cellcolor{samulightgrey}	& \cellcolor{samulightgrey}	& \cellcolor{samulightgrey}	& \cellcolor{lightsalmon}p+	 \\ \cline{2-10}
& anc.	& \color{trolleygrey} i.s.	& \color{trolleygrey} i.s.	& \cellcolor{greyishcyan} s--	& \cellcolor{greyishcyan}w--	& \cellcolor{samulightgrey}	& \cellcolor{samulightgrey}	& \cellcolor{samulightgrey}	& \color{trolleygrey} i.s.	 \\ \hline
\parbox[t]{2mm}{\multirow{4}{*}{\rotatebox[origin=c]{90}{target}}}
& child.	& \color{trolleygrey} i.s.	& \color{trolleygrey} i.s.	& \color{trolleygrey} i.s.	& \color{trolleygrey} i.s.	& \cellcolor{samulightgrey}	& \cellcolor{samulightgrey}	& \cellcolor{samulightgrey}	& \color{trolleygrey} i.s.	 \\ \cline{2-10}
& par.	& \color{trolleygrey} i.s.	& \color{trolleygrey} i.s.	& \color{trolleygrey} i.s.	& \color{trolleygrey} i.s.	& \cellcolor{samulightgrey}	& \cellcolor{samulightgrey}	& \cellcolor{samulightgrey}	& \color{trolleygrey} i.s.	 \\ \cline{2-10}
& desc.	& \color{trolleygrey} i.s.	& \color{trolleygrey} i.s.	& \color{trolleygrey} i.s.	& \color{trolleygrey} i.s.	& \cellcolor{samulightgrey}	& \cellcolor{samulightgrey}	& \cellcolor{samulightgrey}	& \color{trolleygrey} i.s.	 \\ \cline{2-10}
& anc.	& s0	& s0	& s0	& \color{trolleygrey} i.s.	& \cellcolor{samulightgrey}	& \cellcolor{samulightgrey}	& \cellcolor{samulightgrey}	& s0	 \\ \hline
\end{tabular}
\caption{{\bf Summary of the results for hierarchy A.} The columns of the table display different link types, while the rows correspond to the examined node property on either the source (top 4 rows) or the target (bottom 4 rows). The 3$^{\rm rd}$, 4$^{\rm th}$ and 5$^{\rm th}$ columns correspond to forbidden link types highlighted in grey. Symbols inside the cells refer to the following abbreviations: 's+', 's0' and 's-' for indication of strong preference, no preference (neutrality) and strong anti-preference, 'w+' and 'w-' for weak preference and anti-preference, while 'p+' and 'p-' symbolize preference or anti-preference with a non-trivial peak, and 'i.s' for insufficient statistics. }
\label{tab:Apreftab}
\end{table}

\begin{table}[h]
\newcolumntype{?}{!{\vrule width2pt}}
\begin{tabular}{|c|c?c|c|c|c|c|c|c|c|}
\hline
\multicolumn{2}{|c?}{\multirow{4}{*}{B}} & \multicolumn{4}{c|}{link: add}   & \multicolumn{4}{c|}{link: del} \\ \cline{3-10}
\multicolumn{2}{|c?}{}  &  \multicolumn{2}{c|}{source: new} & \multicolumn{2}{c|}{source: old} & \multicolumn{2}{c|}{source: new} & \multicolumn{2}{c|}{source: old} \\ \cline{3-10}
\multicolumn{2}{|c?}{}    & target:  &   target:   &   target: &    target: &  target:  &   target: &  target:  & target:\\
\multicolumn{2}{|c?}{}     & new           &  old        &     new     &    old       &  new   &   old        &  new        &   old  \\ \thickhline
\parbox[t]{2mm}{\multirow{4}{*}{\rotatebox[origin=c]{90}{source}}}
& child.	& \color{trolleygrey} i.s.	& \color{trolleygrey} i.s.	& \cellcolor{lightsalmon} s+	& \color{trolleygrey} i.s.	& \cellcolor{samulightgrey}	& \cellcolor{samulightgrey}	& \cellcolor{samulightgrey}	& \cellcolor{lightsalmon}w+	 \\ \cline{2-10}
& par.	& \color{trolleygrey} i.s.	& \color{trolleygrey} i.s.	& \color{trolleygrey} i.s.	& \color{trolleygrey} i.s.	& \cellcolor{samulightgrey}	& \cellcolor{samulightgrey}	& \cellcolor{samulightgrey}	& \color{trolleygrey} i.s.	 \\ \cline{2-10}
& desc.	& \color{trolleygrey} i.s.	& \color{trolleygrey} i.s.	& \cellcolor{lightsalmon} s+	& \color{trolleygrey} i.s.	& \cellcolor{samulightgrey}	& \cellcolor{samulightgrey}	& \cellcolor{samulightgrey}	& \cellcolor{lightsalmon}w+	 \\ \cline{2-10}
& anc.	& \cellcolor{greyishcyan}w--	& \color{trolleygrey} i.s.	& \cellcolor{greyishcyan} p--	& \color{trolleygrey} i.s.	& \cellcolor{samulightgrey}	& \cellcolor{samulightgrey}	& \cellcolor{samulightgrey}	& \cellcolor{greyishcyan}w--	 \\ \hline
\parbox[t]{2mm}{\multirow{4}{*}{\rotatebox[origin=c]{90}{target}}}
& child.	& \color{trolleygrey} i.s.	& \color{trolleygrey} i.s.	& \color{trolleygrey} i.s.	& \color{trolleygrey} i.s.	& \cellcolor{samulightgrey}	& \cellcolor{samulightgrey}	& \cellcolor{samulightgrey}	& \color{trolleygrey} i.s.	 \\ \cline{2-10}
& par.	& \color{trolleygrey} i.s.	& \color{trolleygrey} i.s.	& \color{trolleygrey} i.s.	& \color{trolleygrey} i.s.	& \cellcolor{samulightgrey}	& \cellcolor{samulightgrey}	& \cellcolor{samulightgrey}	& \color{trolleygrey} i.s.	 \\ \cline{2-10}
& desc.	& \color{trolleygrey} i.s.	& \color{trolleygrey} i.s.	& \color{trolleygrey} i.s.	& \color{trolleygrey} i.s.	& \cellcolor{samulightgrey}	& \cellcolor{samulightgrey}	& \cellcolor{samulightgrey}	& \color{trolleygrey} i.s.	 \\ \cline{2-10}
& anc.	& s0	& s0	& s0	& \color{trolleygrey} i.s.	& \cellcolor{samulightgrey}	& \cellcolor{samulightgrey}	& \cellcolor{samulightgrey}	& \cellcolor{lightsalmon} p+	 \\ \hline
\end{tabular}
\caption{{\bf Summary of the results for hierarchy B.} The arrangement and abbreviations of the table are the same as in case of \ref{tab:Apreftab} Table.
}
\label{tab:Bpreftab}
\end{table}

\begin{table}[h]
\newcolumntype{?}{!{\vrule width2pt}}
\begin{tabular}{|c|c?c|c|c|c|c|c|c|c|}
\hline
\multicolumn{2}{|c?}{\multirow{4}{*}{C}} & \multicolumn{4}{c|}{link: add}   & \multicolumn{4}{c|}{link: del} \\ \cline{3-10}
\multicolumn{2}{|c?}{}  &  \multicolumn{2}{c|}{source: new} & \multicolumn{2}{c|}{source: old} & \multicolumn{2}{c|}{source: new} & \multicolumn{2}{c|}{source: old} \\ \cline{3-10}
\multicolumn{2}{|c?}{}    & target:  &   target:   &   target: &    target: &  target:  &   target: &  target:  & target:\\
\multicolumn{2}{|c?}{}     & new           &  old        &     new     &    old       &  new   &   old        &  new        &   old  \\ \thickhline
\parbox[t]{2mm}{\multirow{4}{*}{\rotatebox[origin=c]{90}{source}}}
& child.	& \color{trolleygrey} i.s.	& \color{trolleygrey} i.s.	& \cellcolor{lightsalmon} s+	& \cellcolor{lightsalmon} s+	& \cellcolor{samulightgrey}	& \cellcolor{samulightgrey}	& \cellcolor{samulightgrey}	& s0	 \\ \cline{2-10}
& par.	& \color{trolleygrey} i.s.	& \color{trolleygrey} i.s.	& \color{trolleygrey} i.s.	& \color{trolleygrey} i.s.	& \cellcolor{samulightgrey}	& \cellcolor{samulightgrey}	& \cellcolor{samulightgrey}	& \color{trolleygrey} i.s.	 \\ \cline{2-10}
& desc.	& \color{trolleygrey} i.s.	& \cellcolor{lightsalmon} s+	& \cellcolor{lightsalmon} s+	& \cellcolor{lightsalmon} s+	& \cellcolor{samulightgrey}	& \cellcolor{samulightgrey}	& \cellcolor{samulightgrey}	& s0	 \\ \cline{2-10}
& anc.	& s0	& \cellcolor{greyishcyan} s--	& \cellcolor{greyishcyan} s--	& \cellcolor{greyishcyan} s--	& \cellcolor{samulightgrey}	& \cellcolor{samulightgrey}	& \cellcolor{samulightgrey}	& s0	 \\ \hline
\parbox[t]{2mm}{\multirow{4}{*}{\rotatebox[origin=c]{90}{target}}}
& child.	& \color{trolleygrey} i.s.	& \color{trolleygrey} i.s.	& \color{trolleygrey} i.s.	& \color{trolleygrey} i.s.	& \cellcolor{samulightgrey}	& \cellcolor{samulightgrey}	& \cellcolor{samulightgrey}	& \color{trolleygrey} i.s.	 \\ \cline{2-10}
& par.	& \color{trolleygrey} i.s.	& \color{trolleygrey} i.s.	& \color{trolleygrey} i.s.	& \color{trolleygrey} i.s.	& \cellcolor{samulightgrey}	& \cellcolor{samulightgrey}	& \cellcolor{samulightgrey}	& \color{trolleygrey} i.s.	 \\ \cline{2-10}
& desc.	& \color{trolleygrey} i.s.	& \color{trolleygrey} i.s.	& \color{trolleygrey} i.s.	& \color{trolleygrey} i.s.	& \cellcolor{samulightgrey}	& \cellcolor{samulightgrey}	& \cellcolor{samulightgrey}	& \color{trolleygrey} i.s.	 \\ \cline{2-10}
& anc.	& s0	& s0	& \cellcolor{lightsalmon} s+	& s0	& \cellcolor{samulightgrey}	& \cellcolor{samulightgrey}	& \cellcolor{samulightgrey}	& \cellcolor{lightsalmon}w+	 \\ \hline
\end{tabular}
\caption{{\bf Summary of the results for hierarchy C.} The arrangement and abbreviations of the table are the same as in case of \ref{tab:Apreftab} Table.
}
\label{tab:Cpreftab}
\end{table}

\begin{table}[h]
\newcolumntype{?}{!{\vrule width2pt}}
\begin{tabular}{|c|c?c|c|c|c|c|c|c|c|}
\hline
\multicolumn{2}{|c?}{\multirow{4}{*}{D}} & \multicolumn{4}{c|}{link: add}   & \multicolumn{4}{c|}{link: del} \\ \cline{3-10}
\multicolumn{2}{|c?}{}  &  \multicolumn{2}{c|}{source: new} & \multicolumn{2}{c|}{source: old} & \multicolumn{2}{c|}{source: new} & \multicolumn{2}{c|}{source: old} \\ \cline{3-10}
\multicolumn{2}{|c?}{}    & target:  &   target:   &   target: &    target: &  target:  &   target: &  target:  & target:\\
\multicolumn{2}{|c?}{}     & new           &  old        &     new     &    old       &  new   &   old        &  new        &   old  \\ \thickhline
\parbox[t]{2mm}{\multirow{4}{*}{\rotatebox[origin=c]{90}{source}}}
& child.	& \cellcolor{lightsalmon} s+	& \cellcolor{lightsalmon} s+	& \cellcolor{lightsalmon} s+	& \cellcolor{lightsalmon} s+	& \cellcolor{samulightgrey}	& \cellcolor{samulightgrey}	& \cellcolor{samulightgrey}	& \cellcolor{lightsalmon} s+	 \\ \cline{2-10}
& par.	& \color{trolleygrey} i.s.	& \color{trolleygrey} i.s.	& \cellcolor{greyishcyan}w--	& \color{trolleygrey} i.s.	& \cellcolor{samulightgrey}	& \cellcolor{samulightgrey}	& \cellcolor{samulightgrey}	& \color{trolleygrey} i.s.	 \\ \cline{2-10}
& desc.	& \cellcolor{greyishcyan} p--	& \cellcolor{lightsalmon} s+	& \cellcolor{lightsalmon} s+	& \cellcolor{lightsalmon} p+	& \cellcolor{samulightgrey}	& \cellcolor{samulightgrey}	& \cellcolor{samulightgrey}	& \cellcolor{lightsalmon} p+	 \\ \cline{2-10}
& anc.	& s0	& \cellcolor{greyishcyan} s--	& \cellcolor{greyishcyan} s--	& \cellcolor{greyishcyan} s--	& \cellcolor{samulightgrey}	& \cellcolor{samulightgrey}	& \cellcolor{samulightgrey}	& s0	 \\ \hline
\parbox[t]{2mm}{\multirow{4}{*}{\rotatebox[origin=c]{90}{target}}}
& child.	& \color{trolleygrey} i.s.	& \color{trolleygrey} i.s.	& \cellcolor{greyishcyan} s--	& s0	& \cellcolor{samulightgrey}	& \cellcolor{samulightgrey}	& \cellcolor{samulightgrey}	& s0	 \\ \cline{2-10}
& par.	& \color{trolleygrey} i.s.	& \color{trolleygrey} i.s.	& \cellcolor{lightsalmon} s+	& s0	& \cellcolor{samulightgrey}	& \cellcolor{samulightgrey}	& \cellcolor{samulightgrey}	& \cellcolor{lightsalmon} s+	 \\ \cline{2-10}
& desc.	& \cellcolor{greyishcyan}w--	& \color{trolleygrey} i.s.	& \cellcolor{greyishcyan} s--	& s0	& \cellcolor{samulightgrey}	& \cellcolor{samulightgrey}	& \cellcolor{samulightgrey}	& s0	 \\ \cline{2-10}
& anc.	& \cellcolor{lightsalmon} s+	& s0	& \cellcolor{lightsalmon} s+	& s0	& \cellcolor{samulightgrey}	& \cellcolor{samulightgrey}	& \cellcolor{samulightgrey}	& \cellcolor{lightsalmon} p+	 \\ \hline
\end{tabular}
\caption{{\bf Summary of the results for hierarchy D.} The arrangement and abbreviations of the table are the same as in case of \ref{tab:Apreftab} Table.
}
\label{tab:Dpreftab}
\end{table}

\begin{table}[h]
\newcolumntype{?}{!{\vrule width2pt}}
\begin{tabular}{|c|c?c|c|c|c|c|c|c|c|}
\hline
\multicolumn{2}{|c?}{\multirow{4}{*}{E}} & \multicolumn{4}{c|}{link: add}   & \multicolumn{4}{c|}{link: del} \\ \cline{3-10}
\multicolumn{2}{|c?}{}  &  \multicolumn{2}{c|}{source: new} & \multicolumn{2}{c|}{source: old} & \multicolumn{2}{c|}{source: new} & \multicolumn{2}{c|}{source: old} \\ \cline{3-10}
\multicolumn{2}{|c?}{}    & target:  &   target:   &   target: &    target: &  target:  &   target: &  target:  & target:\\
\multicolumn{2}{|c?}{}     & new           &  old        &     new     &    old       &  new   &   old        &  new        &   old  \\ \thickhline
\parbox[t]{2mm}{\multirow{4}{*}{\rotatebox[origin=c]{90}{source}}}
& child.	& \color{trolleygrey} i.s.	& \color{trolleygrey} i.s.	& \cellcolor{lightsalmon} s+	& \color{trolleygrey} i.s.	& \cellcolor{samulightgrey}	& \cellcolor{samulightgrey}	& \cellcolor{samulightgrey}	& \cellcolor{lightsalmon}w+	 \\ \cline{2-10}
& par.	& \color{trolleygrey} i.s.	& \color{trolleygrey} i.s.	& \color{trolleygrey} i.s.	& \color{trolleygrey} i.s.	& \cellcolor{samulightgrey}	& \cellcolor{samulightgrey}	& \cellcolor{samulightgrey}	& \color{trolleygrey} i.s.	 \\ \cline{2-10}
& desc.	& \color{trolleygrey} i.s.	& \color{trolleygrey} i.s.	& \cellcolor{lightsalmon} s+	& \color{trolleygrey} i.s.	& \cellcolor{samulightgrey}	& \cellcolor{samulightgrey}	& \cellcolor{samulightgrey}	& \cellcolor{lightsalmon}w+	 \\ \cline{2-10}
& anc.	& \cellcolor{greyishcyan}w--	& \color{trolleygrey} i.s.	& \cellcolor{greyishcyan} s--	& \cellcolor{greyishcyan} s--	& \cellcolor{samulightgrey}	& \cellcolor{samulightgrey}	& \cellcolor{samulightgrey}	& s0	 \\ \hline
\parbox[t]{2mm}{\multirow{4}{*}{\rotatebox[origin=c]{90}{target}}}
& child.	& \color{trolleygrey} i.s.	& \color{trolleygrey} i.s.	& \color{trolleygrey} i.s.	& \color{trolleygrey} i.s.	& \cellcolor{samulightgrey}	& \cellcolor{samulightgrey}	& \cellcolor{samulightgrey}	& \color{trolleygrey} i.s.	 \\ \cline{2-10}
& par.	& \color{trolleygrey} i.s.	& \color{trolleygrey} i.s.	& \color{trolleygrey} i.s.	& \color{trolleygrey} i.s.	& \cellcolor{samulightgrey}	& \cellcolor{samulightgrey}	& \cellcolor{samulightgrey}	& \color{trolleygrey} i.s.	 \\ \cline{2-10}
& desc.	& \color{trolleygrey} i.s.	& \color{trolleygrey} i.s.	& \color{trolleygrey} i.s.	& \color{trolleygrey} i.s.	& \cellcolor{samulightgrey}	& \cellcolor{samulightgrey}	& \cellcolor{samulightgrey}	& \color{trolleygrey} i.s.	 \\ \cline{2-10}
& anc.	& s0	& \color{trolleygrey} i.s.	& s0	& s0	& \cellcolor{samulightgrey}	& \cellcolor{samulightgrey}	& \cellcolor{samulightgrey}	& s0	 \\ \hline
\end{tabular}
\caption{{\bf Summary of the results for hierarchy E.} The arrangement and abbreviations of the table are the same as in case of \ref{tab:Apreftab} Table.
}
\label{tab:Epreftab}
\end{table}

\begin{table}[h]
\newcolumntype{?}{!{\vrule width2pt}}
\begin{tabular}{|c|c?c|c|c|c|c|c|c|c|}
\hline
\multicolumn{2}{|c?}{\multirow{4}{*}{G}} & \multicolumn{4}{c|}{link: add}   & \multicolumn{4}{c|}{link: del} \\ \cline{3-10}
\multicolumn{2}{|c?}{}  &  \multicolumn{2}{c|}{source: new} & \multicolumn{2}{c|}{source: old} & \multicolumn{2}{c|}{source: new} & \multicolumn{2}{c|}{source: old} \\ \cline{3-10}
\multicolumn{2}{|c?}{}    & target:  &   target:   &   target: &    target: &  target:  &   target: &  target:  & target:\\
\multicolumn{2}{|c?}{}     & new           &  old        &     new     &    old       &  new   &   old        &  new        &   old  \\ \thickhline
\parbox[t]{2mm}{\multirow{4}{*}{\rotatebox[origin=c]{90}{source}}}
& child.	& \color{trolleygrey} i.s.	& \color{trolleygrey} i.s.	& \cellcolor{lightsalmon} s+	& \cellcolor{lightsalmon} s+	& \cellcolor{samulightgrey}	& \cellcolor{samulightgrey}	& \cellcolor{samulightgrey}	& \cellcolor{lightsalmon} s+	 \\ \cline{2-10}
& par.	& \color{trolleygrey} i.s.	& \color{trolleygrey} i.s.	& \color{trolleygrey} i.s.	& \color{trolleygrey} i.s.	& \cellcolor{samulightgrey}	& \cellcolor{samulightgrey}	& \cellcolor{samulightgrey}	& \color{trolleygrey} i.s.	 \\ \cline{2-10}
& desc.	& \color{trolleygrey} i.s.	& \color{trolleygrey} i.s.	& \cellcolor{lightsalmon} p+	& \cellcolor{lightsalmon} s+	& \cellcolor{samulightgrey}	& \cellcolor{samulightgrey}	& \cellcolor{samulightgrey}	& \cellcolor{lightsalmon} s+	 \\ \cline{2-10}
& anc.	& \cellcolor{greyishcyan}w--	& \cellcolor{greyishcyan} s--	& \cellcolor{greyishcyan} p--	& \cellcolor{greyishcyan} s--	& \cellcolor{samulightgrey}	& \cellcolor{samulightgrey}	& \cellcolor{samulightgrey}	& \cellcolor{greyishcyan} s--	 \\ \hline
\parbox[t]{2mm}{\multirow{4}{*}{\rotatebox[origin=c]{90}{target}}}
& child.	& \color{trolleygrey} i.s.	& \color{trolleygrey} i.s.	& \color{trolleygrey} i.s.	& \color{trolleygrey} i.s.	& \cellcolor{samulightgrey}	& \cellcolor{samulightgrey}	& \cellcolor{samulightgrey}	& \cellcolor{lightsalmon}w+	 \\ \cline{2-10}
& par.	& \color{trolleygrey} i.s.	& \color{trolleygrey} i.s.	& \color{trolleygrey} i.s.	& \color{trolleygrey} i.s.	& \cellcolor{samulightgrey}	& \cellcolor{samulightgrey}	& \cellcolor{samulightgrey}	& \color{trolleygrey} i.s.	 \\ \cline{2-10}
& desc.	& \color{trolleygrey} i.s.	& \color{trolleygrey} i.s.	& \color{trolleygrey} i.s.	& \color{trolleygrey} i.s.	& \cellcolor{samulightgrey}	& \cellcolor{samulightgrey}	& \cellcolor{samulightgrey}	& \cellcolor{lightsalmon}w+	 \\ \cline{2-10}
& anc.	& s0	& \cellcolor{greyishcyan}w--	& s0	& s0	& \cellcolor{samulightgrey}	& \cellcolor{samulightgrey}	& \cellcolor{samulightgrey}	& s0	 \\ \hline
\end{tabular}
\caption{{\bf Summary of the results for hierarchy G.} The arrangement and abbreviations of the table are the same as in case of \ref{tab:Apreftab} Table.
}
\label{tab:Gpreftab}
\end{table}

\begin{table}[h]
\newcolumntype{?}{!{\vrule width2pt}}
\begin{tabular}{|c|c?c|c|c|c|c|c|c|c|}
\hline
\multicolumn{2}{|c?}{\multirow{4}{*}{N}} & \multicolumn{4}{c|}{link: add}   & \multicolumn{4}{c|}{link: del} \\ \cline{3-10}
\multicolumn{2}{|c?}{}  &  \multicolumn{2}{c|}{source: new} & \multicolumn{2}{c|}{source: old} & \multicolumn{2}{c|}{source: new} & \multicolumn{2}{c|}{source: old} \\ \cline{3-10}
\multicolumn{2}{|c?}{}    & target:  &   target:   &   target: &    target: &  target:  &   target: &  target:  & target:\\
\multicolumn{2}{|c?}{}     & new           &  old        &     new     &    old       &  new   &   old        &  new        &   old  \\ \thickhline
\parbox[t]{2mm}{\multirow{4}{*}{\rotatebox[origin=c]{90}{source}}}
& child.	& \color{trolleygrey} i.s.	& \color{trolleygrey} i.s.	& \color{trolleygrey} i.s.	& \color{trolleygrey} i.s.	& \cellcolor{samulightgrey}	& \cellcolor{samulightgrey}	& \cellcolor{samulightgrey}	& \color{trolleygrey} i.s.	 \\ \cline{2-10}
& par.	& \color{trolleygrey} i.s.	& \color{trolleygrey} i.s.	& \color{trolleygrey} i.s.	& \color{trolleygrey} i.s.	& \cellcolor{samulightgrey}	& \cellcolor{samulightgrey}	& \cellcolor{samulightgrey}	& \color{trolleygrey} i.s.	 \\ \cline{2-10}
& desc.	& \color{trolleygrey} i.s.	& \color{trolleygrey} i.s.	& \color{trolleygrey} i.s.	& \cellcolor{lightsalmon}w+	& \cellcolor{samulightgrey}	& \cellcolor{samulightgrey}	& \cellcolor{samulightgrey}	& \color{trolleygrey} i.s.	 \\ \cline{2-10}
& anc.	& \cellcolor{greyishcyan}w--	& s0	& \cellcolor{greyishcyan} s--	& s0	& \cellcolor{samulightgrey}	& \cellcolor{samulightgrey}	& \cellcolor{samulightgrey}	& s0	 \\ \hline
\parbox[t]{2mm}{\multirow{4}{*}{\rotatebox[origin=c]{90}{target}}}
& child.	& \color{trolleygrey} i.s.	& \color{trolleygrey} i.s.	& \color{trolleygrey} i.s.	& \color{trolleygrey} i.s.	& \cellcolor{samulightgrey}	& \cellcolor{samulightgrey}	& \cellcolor{samulightgrey}	& \color{trolleygrey} i.s.	 \\ \cline{2-10}
& par.	& \color{trolleygrey} i.s.	& \color{trolleygrey} i.s.	& \color{trolleygrey} i.s.	& \color{trolleygrey} i.s.	& \cellcolor{samulightgrey}	& \cellcolor{samulightgrey}	& \cellcolor{samulightgrey}	& \color{trolleygrey} i.s.	 \\ \cline{2-10}
& desc.	& \color{trolleygrey} i.s.	& \color{trolleygrey} i.s.	& \color{trolleygrey} i.s.	& \color{trolleygrey} i.s.	& \cellcolor{samulightgrey}	& \cellcolor{samulightgrey}	& \cellcolor{samulightgrey}	& \color{trolleygrey} i.s.	 \\ \cline{2-10}
& anc.	& \cellcolor{lightsalmon}w+	& s0	& s0	& s0	& \cellcolor{samulightgrey}	& \cellcolor{samulightgrey}	& \cellcolor{samulightgrey}	& s0	 \\ \hline
\end{tabular}
\caption{{\bf Summary of the results for hierarchy N.} The arrangement and abbreviations of the table are the same as in case of \ref{tab:Apreftab} Table.
}
\label{tab:Npreftab}
\end{table}

\section{Measuring preference in the attachment and detachment events}

\begin{figure}[h]
\centering
\includegraphics[width=1.\textwidth]{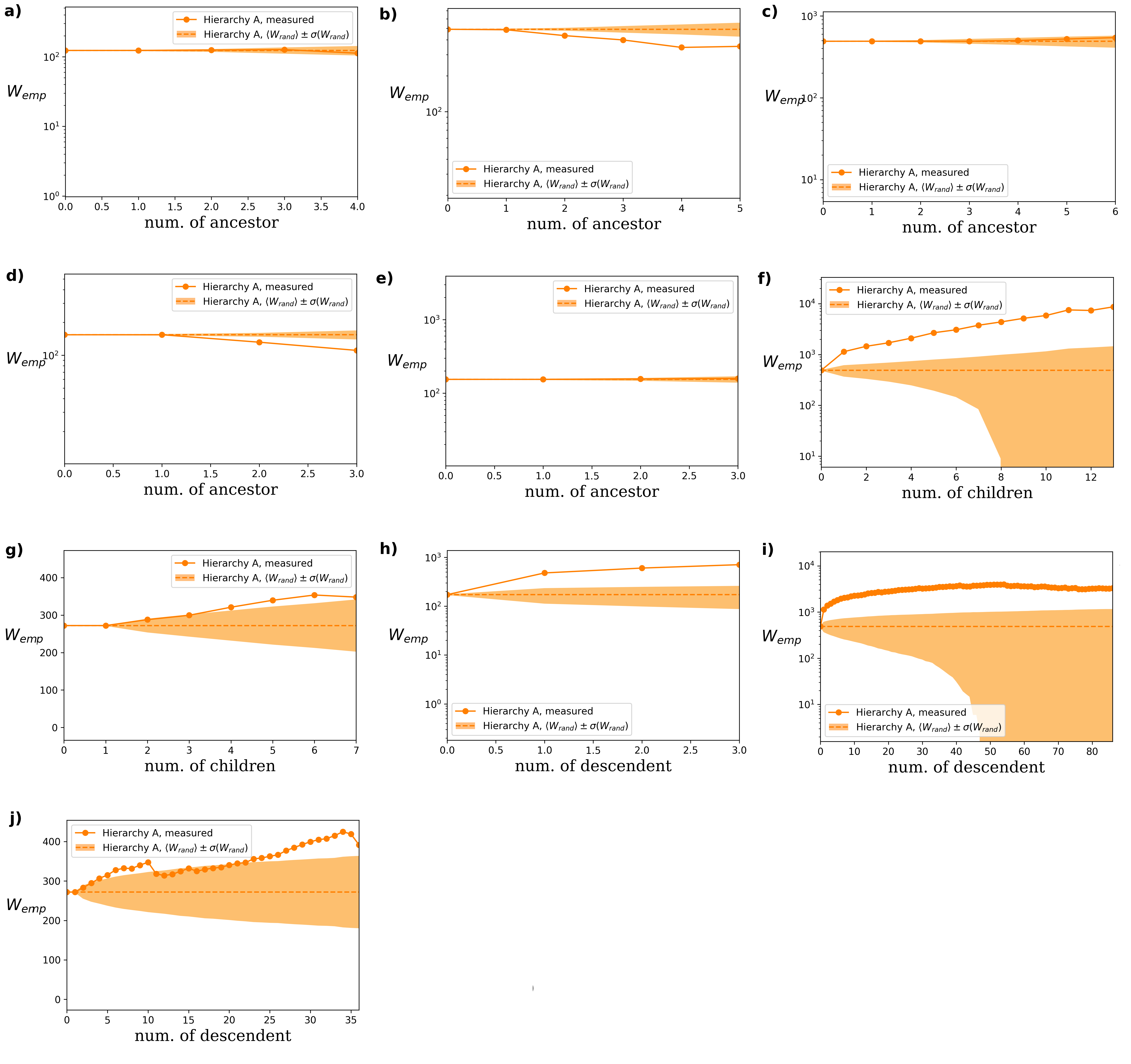}
\caption{{\bf Results for hierarchy A.} a) Addition of new links between new nodes, on the horizontal axis we show the num. of ancestors of the target node. b) Addition of new links pointing from old to new nodes, on the horizontal axis we show the num. of ancestors of the source node. c) Addition of new links pointing from old to new nodes, on the horizontal axis we show the num. of ancestors of the target node. d) Addition of new links between old nodes, on the horizontal axis we show the num. of ancestors of the source node. e) Addition of new links between old nodes, on the horizontal axis we show the num. of ancestors of the target node. f) Addition of new links pointing from old to new nodes, on the horizontal axis we show the num. of children of the source node. g) Deletion of links between old nodes, on the horizontal axis we show the num. of children of the source node. h) Addition of new links pointing from new to old nodes, on the horizontal axis we show the num. of descendents of the source node. i) Addition of new links pointing from old to new nodes, on the horizontal axis we show the num. of descendents of the source node. j) Deletion of links between old nodes, on the horizontal axis we show the num. of descendents of the source node.
}
\label{fig:Aaggr}
\end{figure}

\begin{figure}[h]
\centering
\includegraphics[width=1.\textwidth]{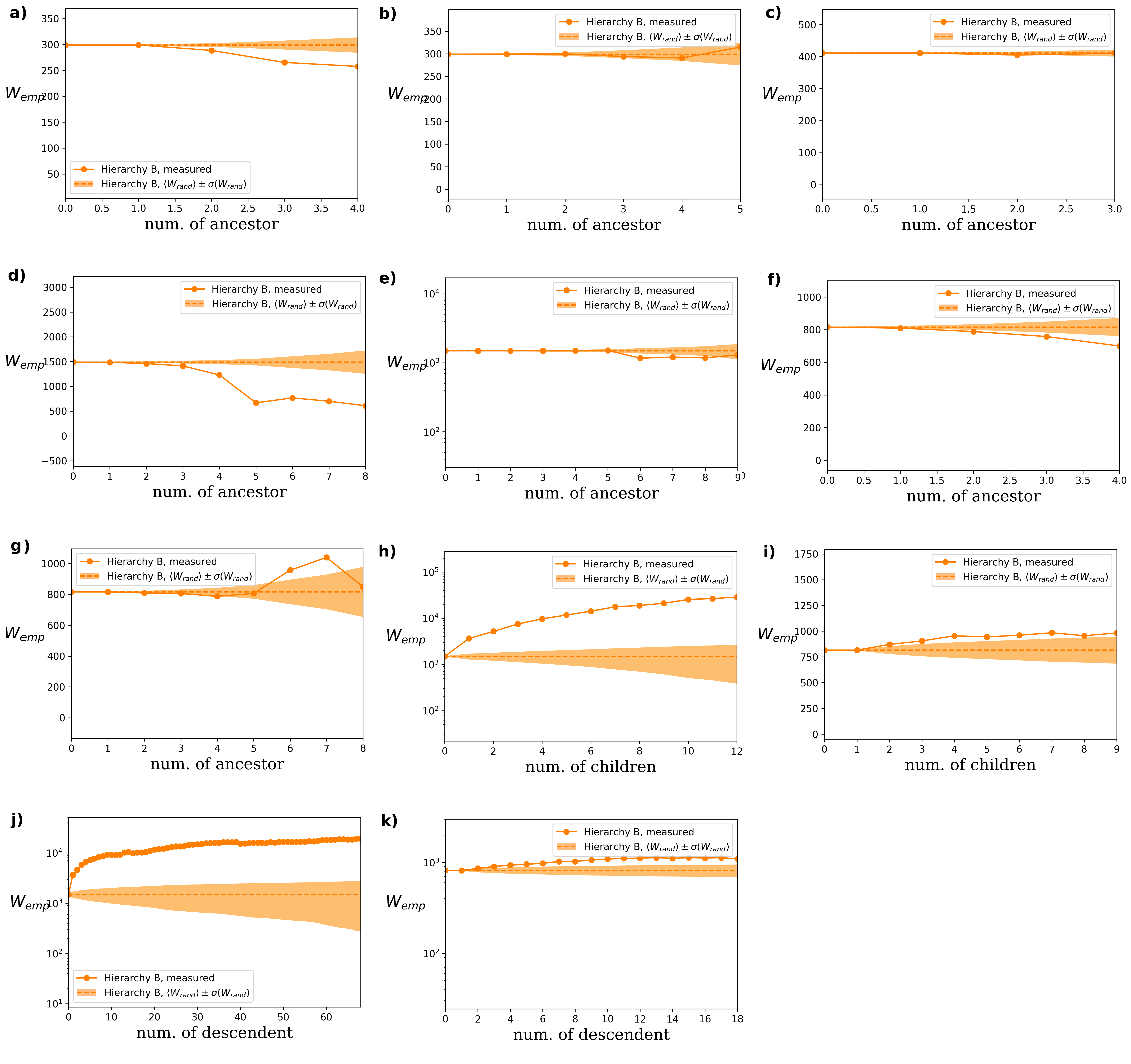}
\caption{{\bf Results for hierarchy B.} 
a) Addition of new links between new nodes, on the horizontal axis we show the num. of ancestors of the source node. b) Addition of new links between new nodes, on the horizontal axis we show the num. of ancestors of the target node. c) Addition of new links pointing from new to old nodes, on the horizontal axis we show the num. of ancestors of the target node. d) Addition of new links pointing from old to new nodes, on the horizontal axis we show the num. of ancestors of the source node. e) Addition of new links pointing from old to new nodes, on the horizontal axis we show the num. of ancestors of the target node. f) Deletion of links between old nodes, on the horizontal axis we show the num. of ancestors of the source node. g) Deletion of links between old nodes, on the horizontal axis we show the num. of ancestors of the target node. h) Addition of new links pointing from old to new nodes, on the horizontal axis we show the num. of children of the source node. i) Deletion of links between old nodes, on the horizontal axis we show the num. of children of the source node. j) Addition of new links pointing from old to new nodes, on the horizontal axis we show the num. of descendents of the source node. k) Deletion of links between old nodes, on the horizontal axis we show the num. of descendents of the source node. }
\label{fig:Baggr}
\end{figure}

\begin{figure}[h]
\centering
\includegraphics[width=1.\textwidth]{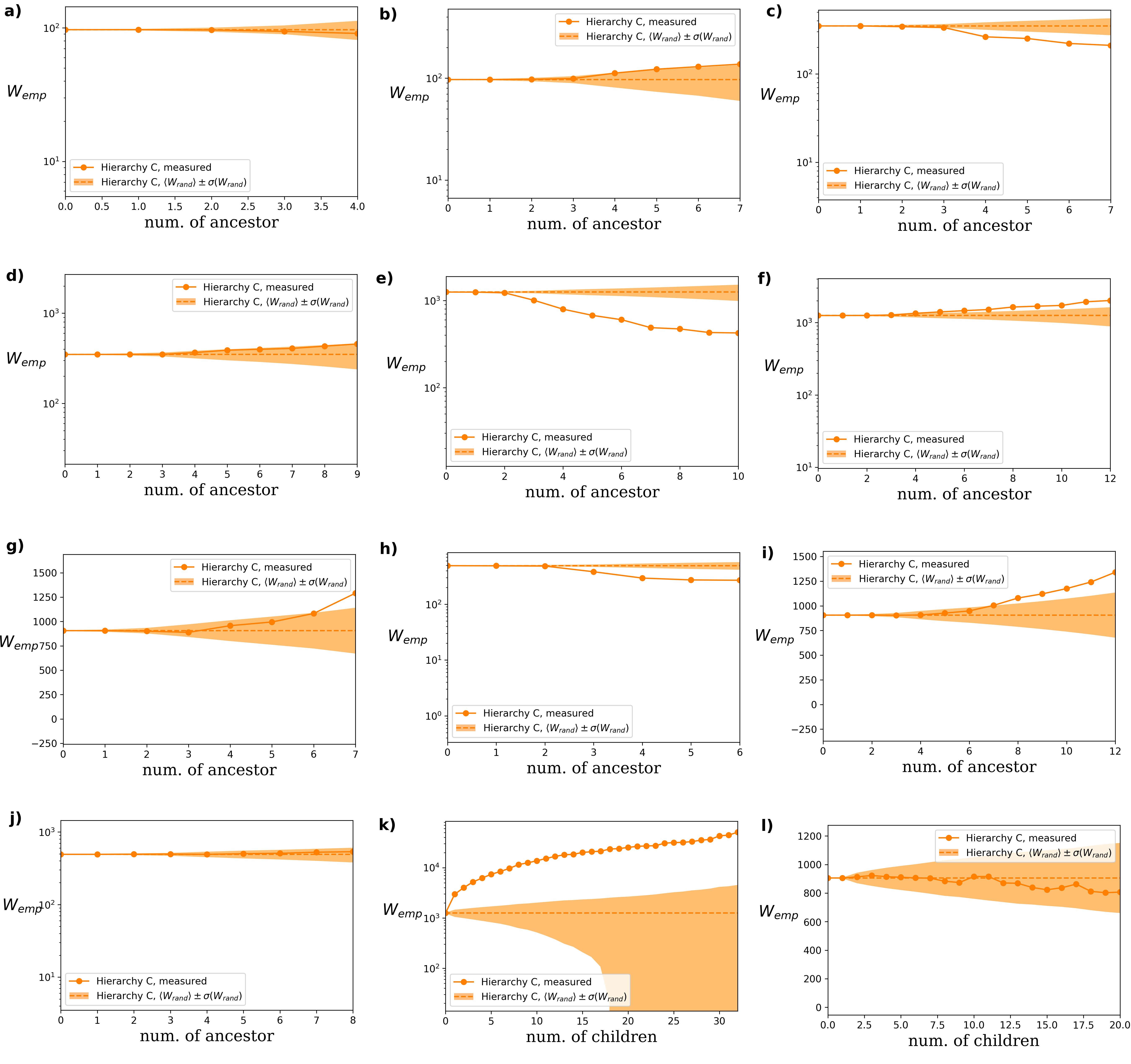}
\caption{{\bf \bf Results for hierarchy C.} 
a) Addition of new links between new nodes, on the horizontal axis we show the num. of ancestors of the source node. 
b) Addition of new links between new nodes, on the horizontal axis we show the num. of ancestors of the target node.
c) Addition of new links pointing from new to old nodes, on the horizontal axis we show the num. of ancestors of the source node. 
d) Addition of new links pointing from new to old nodes, on the horizontal axis we show the num. of ancestors of the target node. 
e) Addition of new links pointing from old to new nodes, on the horizontal axis we show the num. of ancestors of the source node. 
f) Addition of new links pointing from old to new nodes, on the horizontal axis we show the num. of ancestors of the target node. 
g) Deletion of links between old nodes, on the horizontal axis we show the num. of ancestors of the source node.
h) Addition of new links between old nodes, on the horizontal axis we show the num. of ancestors of the source node. 
i) Deletion of links between old nodes, on the horizontal axis we show the num. of ancestors of the target node.
j) Addition of new links between old nodes, on the horizontal axis we show the num. of ancestors of the target node.
k) Addition of new links pointing from old to new nodes, on the horizontal axis we show the num. of children of the source node.
l) Deletion of links between old nodes, on the horizontal axis we show the num. of children of the source node.
}
\label{fig:Caggr}
\end{figure}

\begin{figure}[h]
\centering
\includegraphics[width=1.\textwidth]{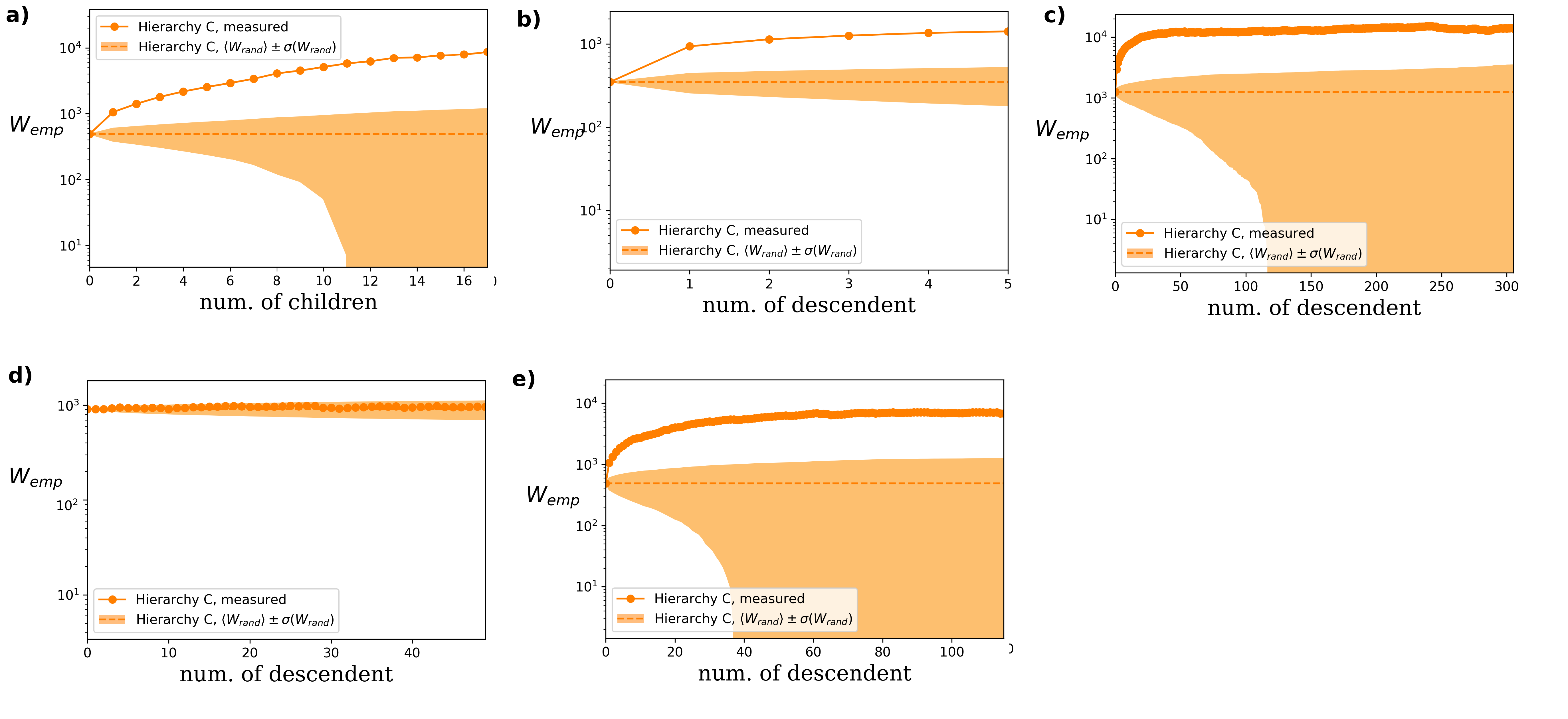}
\caption{{\bf \bf Results for hierarchy C.}
a) Addition of new links between old nodes, on the horizontal axis we show the num. of children of the source node. b) Addition of new links pointing from new to old nodes, on the horizontal axis we show the num. of descendents of the source node. c) Addition of new links pointing from old to new nodes, on the horizontal axis we show the num. of descendents of the source node. d) Deletion of links between old nodes, on the horizontal axis we show the num. of descendents of the source node. e) Addition of new links between old nodes, on the horizontal axis we show the num. of descendents of the source node. 
}
\label{fig:Caggr2}
\end{figure}

\begin{figure}[h]
\centering
\includegraphics[width=1.\textwidth]{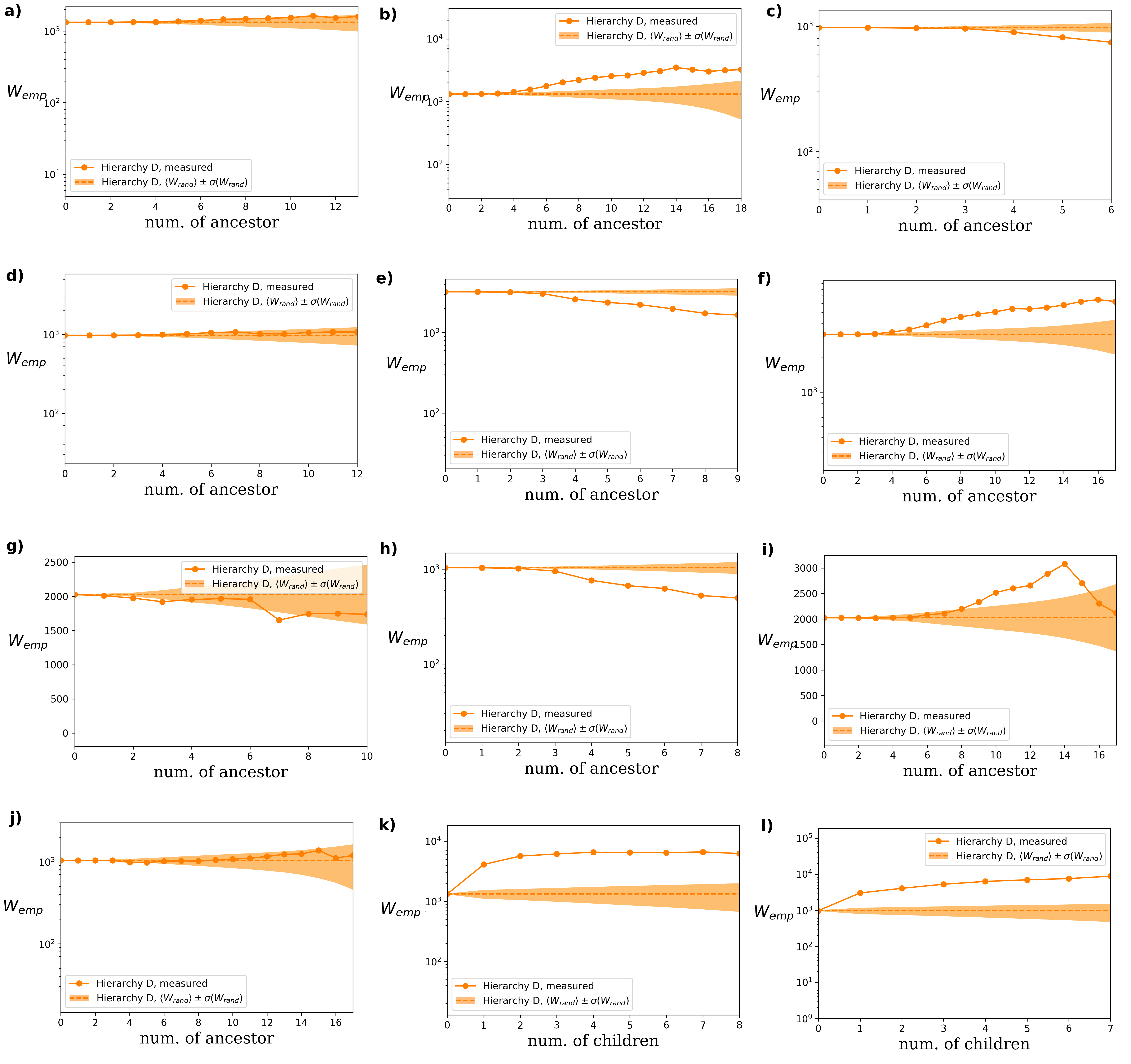}
\caption{{\bf Results for hierarchy D.}
a) Addition of new links between new nodes, on the horizontal axis we show the num. of ancestors of the source node. 
b) Addition of new links between new nodes, on the horizontal axis we show the num. of ancestors of the target node. 
c) Addition of new links pointing from new to old nodes, on the horizontal axis we show the num. of ancestors of the source node. 
d) Addition of new links pointing from new to old nodes, on the horizontal axis we show the num. of ancestors of the target node.
e) Addition of new links pointing from old to new nodes, on the horizontal axis we show the num. of ancestors of the source node. 
f) Addition of new links pointing from old to new nodes, on the horizontal axis we show the num. of ancestors of the target node. 
g) Deletion of links between old nodes, on the horizontal axis we show the num. of ancestors of the source node. 
h) Addition of new links between old nodes, on the horizontal axis we show the num. of ancestors of the source node.
i) Deletion of links between old nodes, on the horizontal axis we show the num. of ancestors of the target node. 
j) Addition of new links between old nodes, on the horizontal axis we show the num. of ancestors of the target node. 
k) Addition of new links between new nodes, on the horizontal axis we show the num. of children of the source node. 
l) Addition of new links pointing from new to old nodes, on the horizontal axis we show the num. of children of the source node.
}
\label{fig:Daggr}
\end{figure}

\begin{figure}[h]
\centering
\includegraphics[width=1.\textwidth]{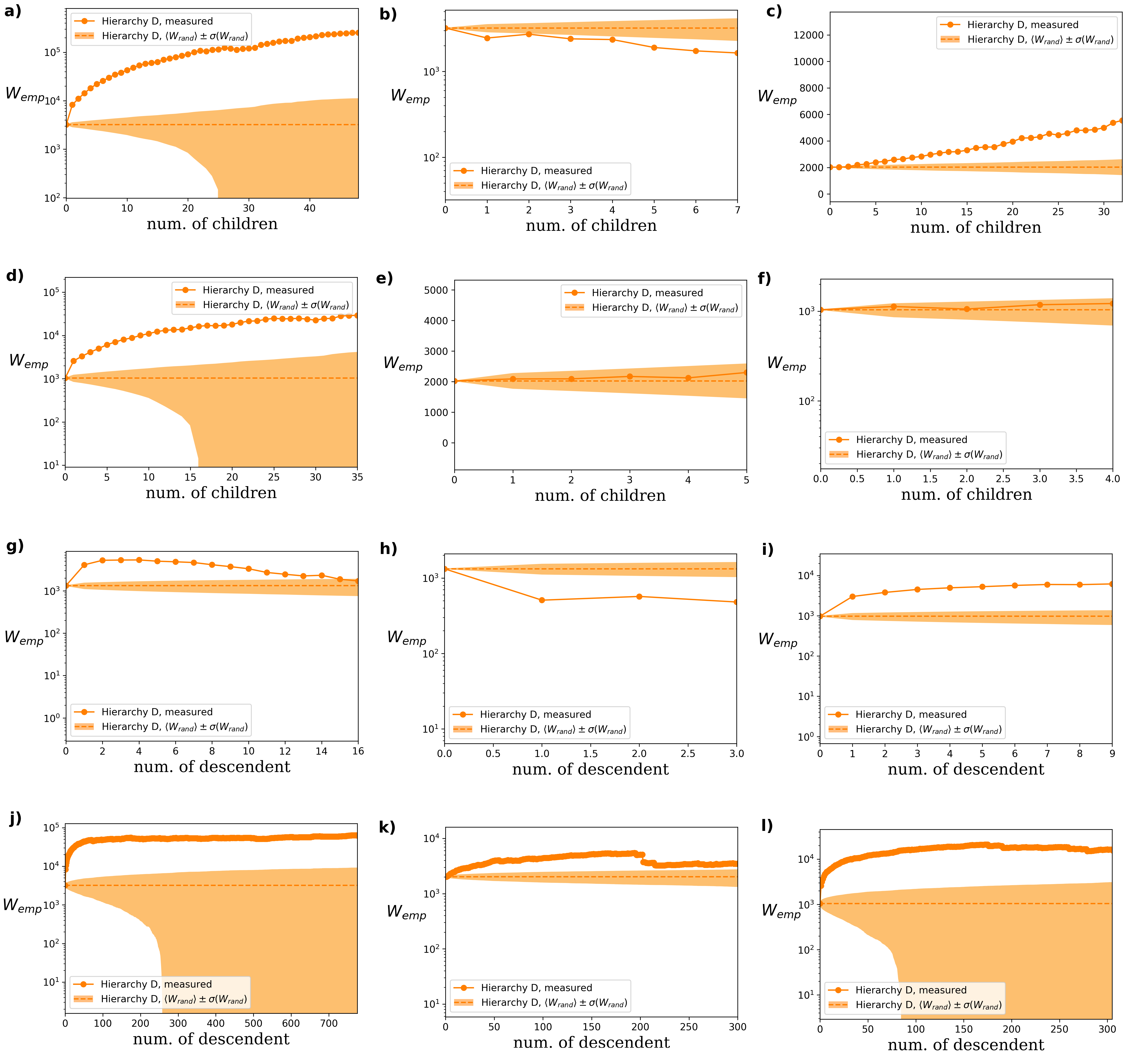}
\caption{{\bf Results for hierarchy D.} 
a) Addition of new links pointing from old to new nodes, on the horizontal axis we show the num. of children of the source node. 
b) Addition of new links pointing from old to new nodes, on the horizontal axis we show the num. of children of the target node. 
c) Deletion of links between old nodes, on the horizontal axis we show the num. of children of the source node. 
d) Addition of new links between old nodes, on the horizontal axis we show the num. of children of the source node. 
e) Deletion of links between old nodes, on the horizontal axis we show the num. of children of the target node. 
f) Addition of new links between old nodes, on the horizontal axis we show the num. of children of the target node. 
g) Addition of new links between new nodes, on the horizontal axis we show the num. of descendents of the source node. 
h) Addition of new links between new nodes, on the horizontal axis we show the num. of descendents of the target node.
i) Addition of new links pointing from new to old nodes, on the horizontal axis we show the num. of descendents of the source node. 
j) Addition of new links pointing from old to new nodes, on the horizontal axis we show the num. of descendents of the source node. 
k) Deletion of links between old nodes, on the horizontal axis we show the num. of descendents of the source node. 
l) Addition of new links between old nodes, on the horizontal axis we show the num. of descendents of the source node. 
}
\label{fig:Daggr2}
\end{figure}

\begin{figure}[h]
\centering
\includegraphics[width=1.\textwidth]{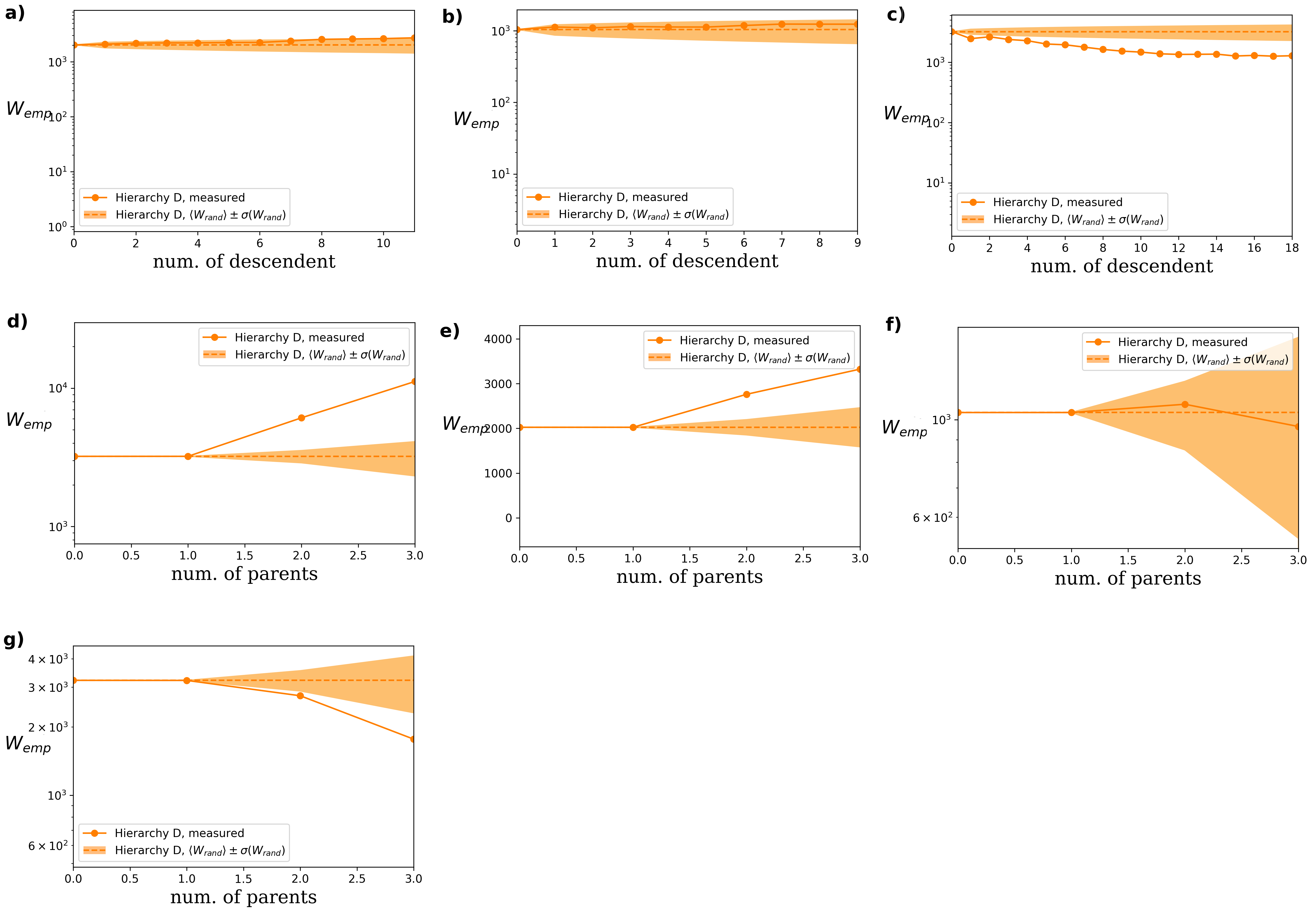}
\caption{{\bf Results for hierarchy D.} 
a) Deletion of links between old nodes, on the horizontal axis we show the num. of descendents of the target node. 
b) Addition of new links between old nodes, on the horizontal axis we show the num. of descendents of the target node. 
c) Addition of new links pointing from old to new nodes, on the horizontal axis we show the num. of descendents of the target node. 
d) Addition of new links pointing from old to new nodes, on the horizontal axis we show the num. of parents of the target node. 
e) Deletion of links between old nodes, on the horizontal axis we show the num. of parents of the target node. 
f) Addition of links between old nodes, on the horizontal axis we show the num. of parents of the target node. 
g) Addition of new links pointing from old to new nodes, on the horizontal axis we show the num. of parents of the source node.
}
\label{fig:Daggr3}
\end{figure}

\begin{figure}[h]
\centering
\includegraphics[width=1.\textwidth]{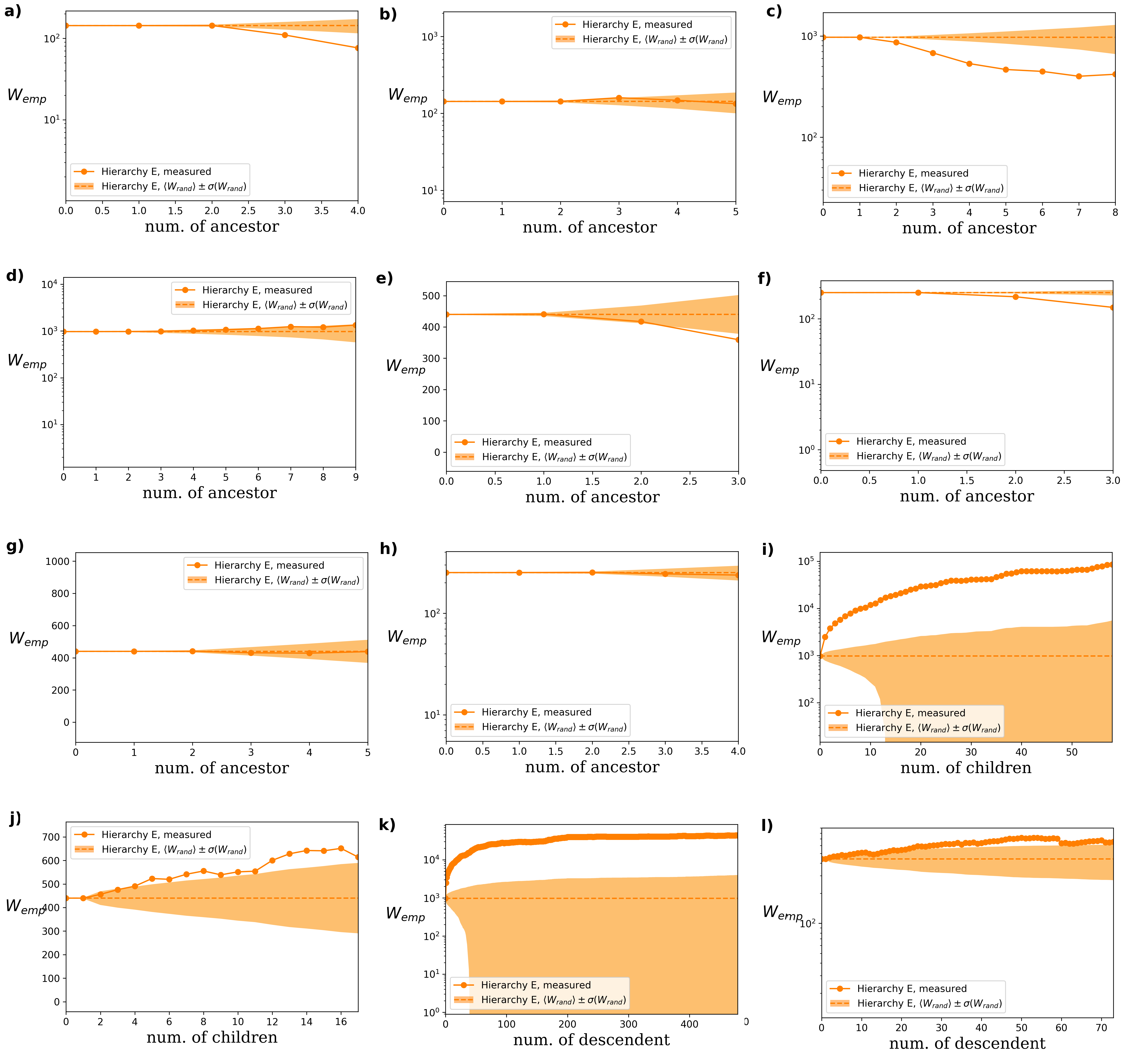}
\caption{{\bf Results for hierarchy E.}
a) Addition of new links between new nodes, on the horizontal axis we show the num. of ancestors of the source node. 
b) Addition of new links between new nodes, on the horizontal axis we show the num. of ancestors of the target node. 
c) Addition of new links pointing from old to new nodes, on the horizontal axis we show the num. of ancestors of the source node. 
d) Addition of new links pointing from old to new nodes, on the horizontal axis we show the num. of ancestors of the target node. 
e) Deletion of links between old nodes, on the horizontal axis we show the num. of ancestors of the source node. 
f) Addition of new between old nodes, on the horizontal axis we show the num. of ancestors of the source node. 
g) Deletion of links between old nodes, on the horizontal axis we show the num. of ancestors of the target node. 
h) Addition of new links between old nodes, on the horizontal axis we show the num. of ancestors of the target node. 
i) Addition of new links pointing from old to new nodes, on the horizontal axis we show the num. of children of the source node. 
j) Deletion of links between old nodes, on the horizontal axis we show the num. of children of the source node. 
k) Addition of new links pointing from old to new nodes, on the horizontal axis we show the num. of descendents of the source node. 
l) Deletion of links between old nodes, on the horizontal axis we show the num. of descendents of the source node.
}
\label{fig:Eaggr}
\end{figure}

\begin{figure}[h]
\centering
\includegraphics[width=1.\textwidth]{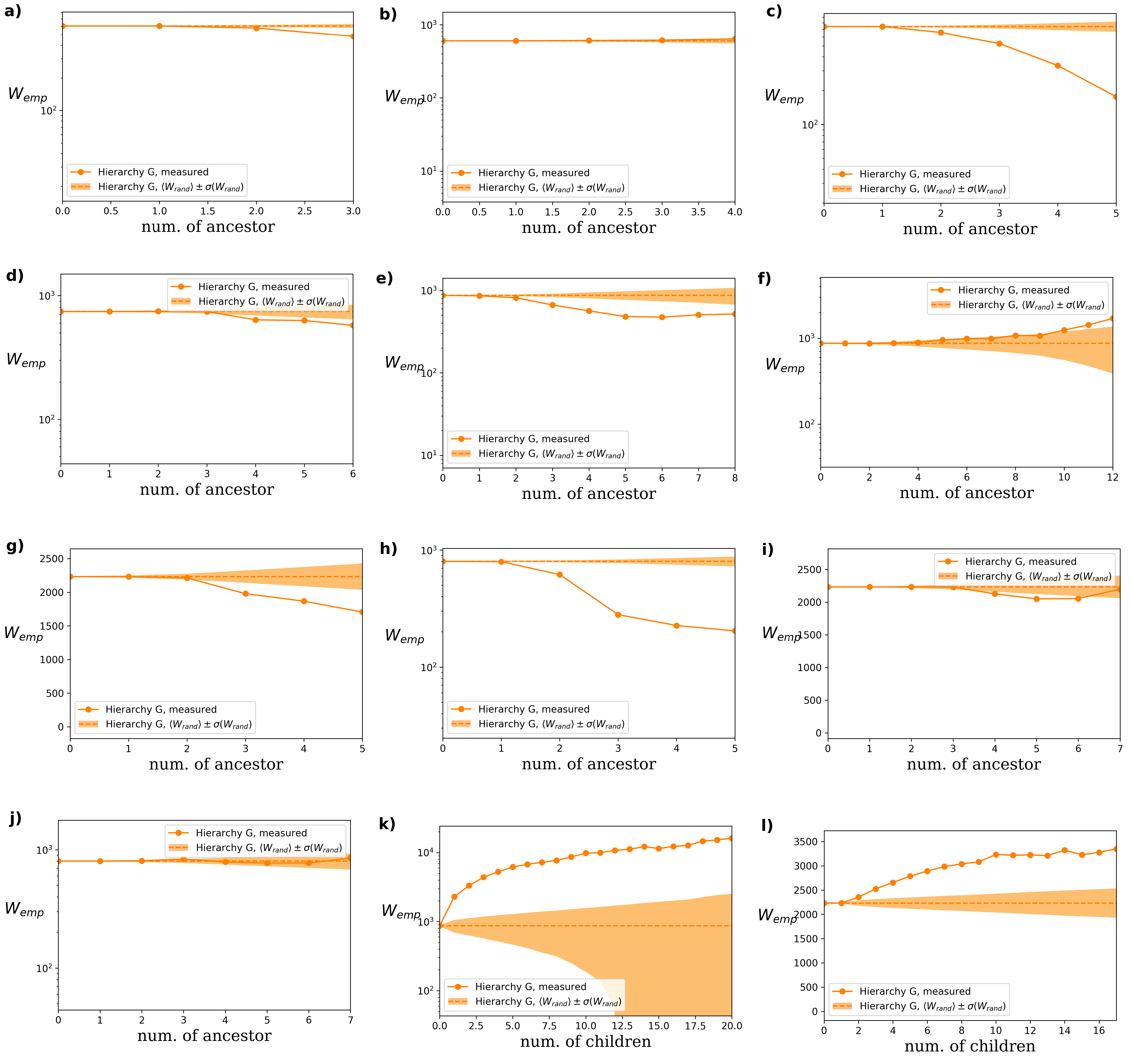}
\caption{{\bf Results for hierarchy G.}
a) Addition of new links between new nodes, on the horizontal axis we show the num. of ancestors of the source node. 
b) Addition of new links between new nodes, on the horizontal axis we show the num. of ancestors of the target node. 
c) Addition of new links pointing from new to old nodes, on the horizontal axis we show the num. of ancestors of the source node. 
d) Addition of new links pointing from new to old nodes, on the horizontal axis we show the num. of ancestors of the target node. 
e) Addition of new links pointing from old to new nodes, on the horizontal axis we show the num. of ancestors of the source node. 
f) Addition of new links pointing from old to new nodes, on the horizontal axis we show the num. of ancestors of the target node. 
g) Deletion of links between old nodes, on the horizontal axis we show the num. of ancestors of the source node. 
h) Addition of new links between old nodes, on the horizontal axis we show the num. of ancestors of the source node.
i) Deletion of links between old nodes, on the horizontal axis we show the num. of ancestors of the target node. 
j) Addition of new links between old nodes, on the horizontal axis we show the num. of ancestors of the target node. 
k) Addition of new links pointing from old to new nodes, on the horizontal axis we show the num. of children of the source node. 
l) Deletion of links between old nodes, on the horizontal axis we show the num. of children of the source node.
}
\label{fig:Gaggr1}
\end{figure}

\begin{figure}[h]
\centering
\includegraphics[width=1.\textwidth]{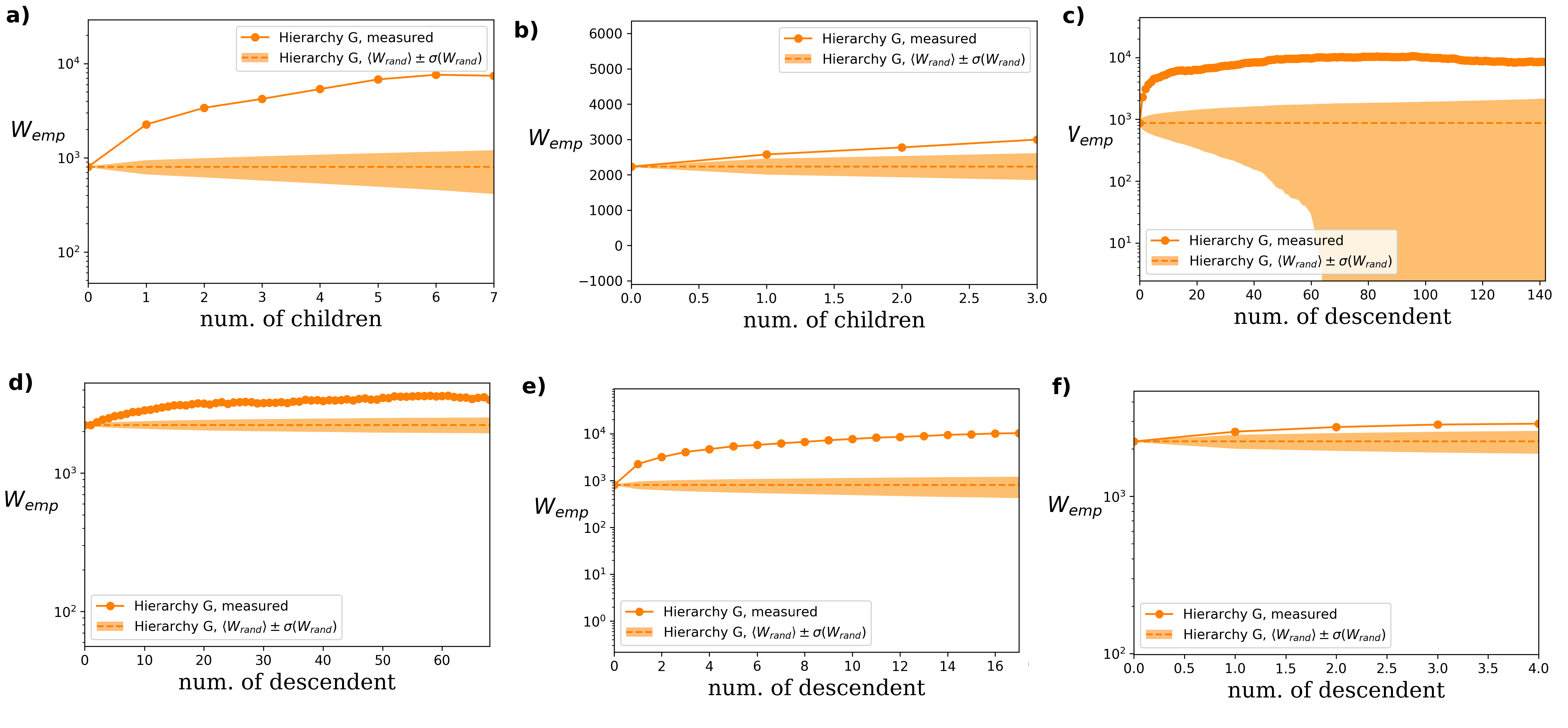}
\caption{{\bf Results for hierarchy G.}
a) Addition of new links between old nodes, on the horizontal axis we show the num. of children of the source node. 
b) Deletion of links between old nodes, on the horizontal axis we show the num. of children of the target node. 
c) Addition of new links pointing from old to new nodes, on the horizontal axis we show the num. of descendents of the source node. 
d) Deletion of links between old nodes, on the horizontal axis we show the num. of descendents of the source node. 
e) Addition of new links between old nodes, on the horizontal axis we show the num. of descendents of the source node. 
f) Deletion of links between old nodes, on the horizontal axis we show the num. of descendents of the target node.
}
\label{fig:Gaggr2}
\end{figure}

\begin{figure}[h]
\centering
\includegraphics[width=1.\textwidth]{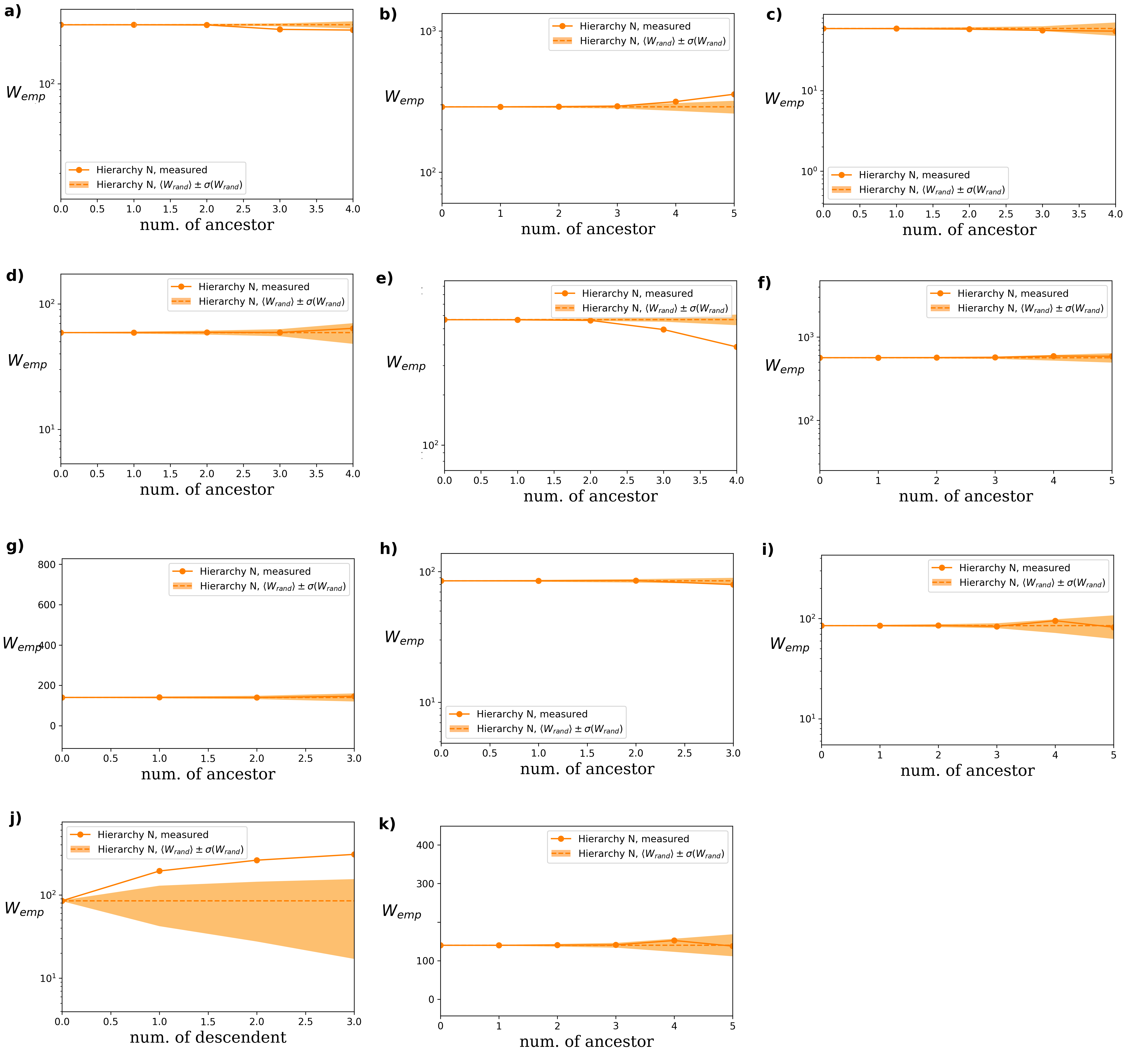}
\caption{{\bf Results for hierarchy N.}
a) Addition of new links between new nodes, on the horizontal axis we show the num. of ancestors of the source node. b) Addition of new links between new nodes, on the horizontal axis we show the num. of ancestors of the target node. c) Addition of new links pointing from new to old nodes, on the horizontal axis we show the num. of ancestors of the source node. d) Addition of new links pointing from new to old nodes, on the horizontal axis we show the num. of ancestors of the target node. e) Addition of new links pointing from old to new nodes, on the horizontal axis we show the num. of ancestors of the source node. f) Addition of new links pointing from old to new nodes, on the horizontal axis we show the num. of ancestors of the target node. g) Deletion of links between old nodes, on the horizontal axis we show the num. of ancestors of the source node. h) Addition of new links between old nodes, on the horizontal axis we show the num. of ancestors of the source node. 
i) Addition of new links between old nodes, on the horizontal axis we show the num. of ancestors of the target node.
j) Addition of links between old nodes, on the horizontal axis we show the num. of descendents of the source node. 
k) Deletion of links between old nodes, on the horizontal axis we show the num. of ancestors of the target node.
}
\label{fig:Naggr}
\end{figure}

The numerical results obtained for hierarchies A,B,C,D,E,G,N are depicted in \ref{fig:Aaggr}-\ref{fig:Naggr} Figs. each of which is composed of several different panels. We display only those attachment/detachment types are in the sub figures where the statistics were found to be sufficient. In each panel $W_{\rm emp}(x)$ defined in Eq.(8) in the main paper as
\begin{equation}
W_{\rm emp}(x)=\sum_{t=1}^{t_{\rm max}-1}\frac{w_{t}(x)}{Q_t(x)}
\end{equation}
is compared to the predicted behavior of the mean and the standard deviation of $W(x)$ (given in Eqs. (9-10) in the main paper) for random events represented by dashed lines in shaded areas.

\end{document}